\documentclass[sigconf, screen]{acmart}

\usepackage{threeparttable}
\usepackage{enumitem}
\usepackage{makecell}
\usepackage[skins]{tcolorbox}
\usepackage{xspace}

\copyrightyear{2026}
\acmYear{2026}
\setcopyright{cc}
\setcctype{by}
\acmConference[MSR '26]{23rd International Conference on Mining Software Repositories}{April 13--14, 2026}{Rio de Janeiro, Brazil}
\acmBooktitle{23rd International Conference on Mining Software Repositories (MSR '26), April 13--14, 2026, Rio de Janeiro, Brazil}
\acmPrice{}
\acmDOI{10.1145/3793302.3793349}
\acmISBN{979-8-4007-2474-9/2026/04}


\tcbset{
  my box2/.style={
    enhanced,
    colframe=#1!80,
    colback=#1!5,
  },
}
\newtcolorbox{summary-rq}{
  my box2=black,
  boxrule=1pt,top=3pt,bottom=3pt,left=4pt,right=4pt
}

\newcommand{\Code}[1]{\begin{small}\texttt{#1}\end{small}}
\newcommand{\numtreated}{806\xspace}

\begin{document}

\title{Speed at the Cost of Quality}
\subtitle{How Cursor AI Increases Short-Term Velocity and Long-Term Complexity in Open-Source Projects}

\author{Hao He}
\orcid{0000-0001-8311-6559}
\affiliation{%
  \institution{Carnegie Mellon University}
  \city{Pittsburgh}
  \state{Pennsylvania}
  \country{USA}
}

\author{Courtney Miller}
\orcid{0000-0002-5297-4523}
\affiliation{%
  \institution{Carnegie Mellon University}
  \city{Pittsburgh}
  \state{Pennsylvania}
  \country{USA}
}

\author{Shyam Agarwal}
\orcid{0009-0009-2147-5674}
\affiliation{%
  \institution{Carnegie Mellon University}
  \city{Pittsburgh}
  \state{Pennsylvania}
  \country{USA}
}

\author{Christian Kästner}
\orcid{0000-0002-4450-4572}
\affiliation{%
  \institution{Carnegie Mellon University}
  \city{Pittsburgh}
  \state{Pennsylvania}
  \country{USA}
}

\author{Bogdan Vasilescu}
\orcid{0000-0003-4418-5783}
\affiliation{%
  \institution{Carnegie Mellon University}
  \city{Pittsburgh}
  \state{Pennsylvania}
  \country{USA}
}

\begin{abstract}
  Large language models (LLMs) have demonstrated the promise to revolutionize the field of software engineering.
  Among other things, LLM agents are rapidly gaining momentum in software development, with practitioners reporting a multifold increase in productivity after adoption.
  Yet, empirical evidence is lacking around these claims.
  In this paper, we estimate the \emph{causal} effect of adopting a widely popular LLM agent assistant, namely Cursor, on \emph{development velocity} and \emph{software quality}.
  The estimation is enabled by a state-of-the-art difference-in-differences design comparing Cursor-adopting GitHub projects with a matched control group of similar GitHub projects that do not use Cursor. 
  We find that the adoption of Cursor leads to a statistically significant, large, but transient increase in project-level development velocity, along with a substantial and persistent increase in static analysis warnings and code complexity.
  Further panel generalized-method-of-moments estimation reveals that increases in static analysis warnings and code complexity are major factors driving long-term velocity slowdown.
  Our study identifies quality assurance as a major bottleneck for early Cursor adopters and calls for it to be a first-class citizen in the design of agentic AI coding tools and AI-driven workflows.
\end{abstract}

\maketitle

\begin{figure}
    \centering
    \includegraphics[width=\linewidth]{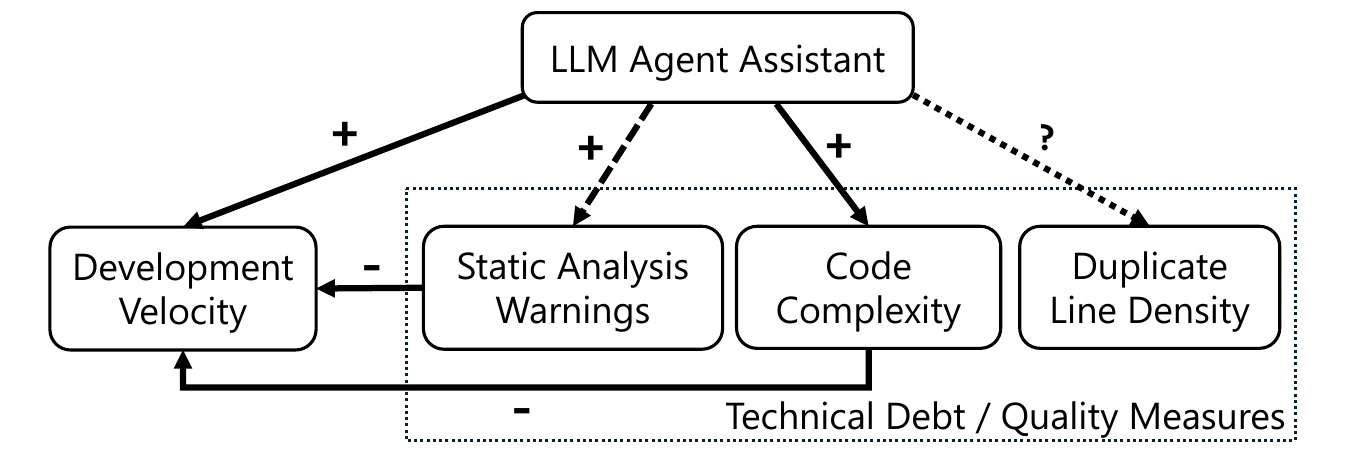}
    \caption{Our theory around how LLM agent assistants impact software development. Solid lines indicate causal relationships supported by our analysis; dashed lines indicate partial evidence; and dotted lines indicate inconclusive evidence.}
    \label{fig:theory}
\end{figure}

\section{Introduction}
\label{sec:intro}

Large language models (LLMs) have demonstrated remarkable capabilities in code generation, achieving near-human performance across various software engineering tasks~\cite{DBLP:conf/fose-ws/FanGHLSYZ23, DBLP:journals/tosem/HouZLYWLLLGW24}. 
Among emerging applications, LLM agent assistants---tools that combine LLMs with autonomous capabilities to inspect project files, execute commands, and iteratively develop code---represent a particularly promising direction for integrating LLMs into software development. 
For example, Cursor~\cite{cursor}, a popular LLM agent assistant, has generated considerable enthusiasm among practitioners, with developers self-reporting multi-fold (as large as 10x) productivity increases and claiming transformative workflow impacts~\cite{cursor-reddit-survey, cursor-10x}.

However, substantial concerns have been raised about the quality of LLM-generated code and the long-term consequences of AI-driven development workflows.
Studies documented that AI coding assistants can produce code with security vulnerabilities~\cite{DBLP:conf/sp/PearceA0DK22,DBLP:conf/ccs/PerryS0B23}, performance issues~\cite{DBLP:conf/issre/LiC0X0S24}, code smells~\cite{DBLP:conf/scam/SiddiqMMJS22}, and increased complexity~\cite{DBLP:journals/tse/LiuTLZZ24}.
Yet, these findings, derived from evaluations of early Codex models in controlled experiments~\cite{DBLP:conf/ccs/PerryS0B23}, completion-based tools like early GitHub Copilot~\cite{DBLP:conf/sp/PearceA0DK22,DBLP:conf/issre/LiC0X0S24,DBLP:conf/scam/SiddiqMMJS22}, or chat-based interfaces like ChatGPT~\cite{DBLP:journals/tse/LiuTLZZ24}, may not generalize to modern LLM agent assistants, which represent a qualitative shift in architecture and integration, not simply incremental improvement.
Completion tools suggest individual lines based on immediate context; chat-based assistants require context-switching to formulate queries and integrate responses.
Both operate at the periphery of the development workflow, with developers remaining the primary agents and maintaining oversight of AI-generated code at a granular level.

LLM agent assistants like Cursor, by contrast, are tightly integrated into the IDE with persistent codebase awareness, autono-mously navigating files, proposing multi-file refactorings, and implementing features spanning dozens of files---all within the development environment.
This architectural difference has profound implications that cannot be extrapolated from prior studies:
Automation scope shifts from accelerating typing to automating entire workflows; seamless integration may enable both productivity gains and over-reliance or reduced code review rigor; quality implications of reviewing large, AI-generated multi-file changes differ fundamentally from reviewing line-by-line completions; and temporal dynamics may involve longer-term effects as technical debt accumulation. 
Consequently, prior findings about security vulnerabilities in Copilot completions or complexity issues in ChatGPT-generated functions provide limited insight into whether---and how---these issues manifest in agentic tools that generate code at a substantially larger scale with different developer oversight patterns.

This gap between generations of AI coding tools also shows up in recent research.
For example, \citet{DBLP:journals/corr/abs-2507-09089} show through controlled experiments that early-2025 AI tools, including Cursor, do not help experienced open-source developers solve real day-to-day tasks faster, pointing to potential slowdown mechanisms such as developer over-optimism, low AI reliability, and high task complexity.
Their findings contradict substantial prior literature on earlier-generation AI coding assistants~\cite{DBLP:conf/pldi/0001KLRRSSA22, DBLP:conf/chi/Vaithilingam0G22, DBLP:conf/icse/Imai22, dohmke2023sea, DBLP:journals/corr/abs-2302-06590, hoffmann2024generative, DBLP:conf/icis/YeverechyahuMO24, DBLP:journals/corr/abs-2410-02091, cui2024effects, DBLP:journals/corr/abs-2406-17910, stray2025developer}, which generally found modest velocity improvements from code completion tools like early GitHub Copilot.

Most recently, \citet{watanabe2025use} take an important step toward understanding modern agentic tools by examining 567 pull requests generated by Claude Code, finding that 83.8\% are accepted and merged.
However, their analysis focuses on PR acceptance rates and task types rather than on longitudinal, project-level effects of tool adoption on development velocity and code quality.
Whether high acceptance rates of individual agent-generated PRs translate into sustained productivity gains and maintained quality at the project level---or whether quality degradation accumulates as teams integrate these tools into everyday workflows over months---remains an open empirical question.
To address this gap, we ask:

\vspace{0.1cm}
\textbf{RQ:} \emph{How does the adoption of LLM agent assistants impact project-level development velocity and code quality?}
\vspace{0.1cm}

We focus on Cursor, one of the earliest and most widely adopted LLM agent assistants~\cite{global-ai-trends}, as our empirical case.
Our key insight is that the presence of \Code{.cursorrules} configuration files in GitHub repositories signals the adoption of Cursor's multi-file editing and agentic capabilities, and thus the date of the first commit touching these configuration files serves as a proxy for the adoption date of the modern Cursor client with agentic coding features.\footnote{
Earlier Cursor versions offered code completion and chat-based code generation.
Cursor rule files were first released in mid-2024, roughly at the same time as the Composer feature, which provides a conversational interface for multi-file code generation.
The agentic features of Composer were released in November 2024 and made default in February 2025~\cite{sarkar2025ai}.
Thus, precisely speaking, our identification captures the adoption of modern Cursor with a mix of Composer's multi-file editing and subsequent transition to the default agentic mode. We call this modern version simply ``Cursor'' for brevity, and we show in the Appendix that limiting our analyses to adoption cohorts before/after agent release/agent made default does not change our main results.} 
By scanning Cursor configuration files across GitHub, we identify \numtreated repositories that adopted the modern Cursor client, with most of the adoptions happening between August 2024 and March 2025.

We then use a \emph{difference-in-differences} (DiD) design with staggered adoption~\cite{callaway2021difference, borusyak2024revisiting}, comparing repositories that adopt Cursor at different times to a matched control group that never adopts during our observation period.
This \emph{quasi-experimental} approach uses naturally occurring variation in adoption timing to identify causal effects while controlling for repository-specific characteristics and common temporal trends.
To construct a comparable control group, we use propensity score matching~\cite{austin2011introduction} to select 1,380 similar repositories from those never adopting Cursor during our observation period.
Our matching model incorporates the dynamic history of repository characteristics---activity levels, contributor counts, and development patterns over the six months preceding potential adoption---ensuring treated and control repositories exhibit similar observable trajectories before adoption.

To estimate treatment effects, we use \emph{the \citet{borusyak2024revisiting} imputation estimator}, a modern DiD approach designed for staggered adoption that avoids biases from traditional two-way fixed effects models.
Using this approach, we estimate the impact of Cursor adoption on two velocity outcomes (commits and lines added) and three code quality outcomes (static analysis warnings, code complexity, and duplicate line density).
Finally, to test temporal interactions between outcomes, we also estimate the impact of changes in code quality on future development velocity (and vice versa) using \emph{panel generalized method of moments} (GMM) models~\cite{arellano1991some}.

Our findings reveal a concerning picture among GitHub open-source projects adopting Cursor.
First, the adoption of Cursor leads to significant, large, but transient velocity increases: Projects experience 3-5x increases in lines added in the first adoption month, but gains dissipate after two months.
Concurrently, we observe persistent technical debt accumulation: Static analysis warnings increase by 30\% and code complexity increases by 41\% post-adoption according to the \citet{borusyak2024revisiting} DiD estimator.
Panel GMM models reveal that accumulated technical debt subsequently reduces future velocity, creating a self-reinforcing cycle.
Notably, Cursor adoption still leads to significant increases in code complexity, even when models control for project velocity dynamics.

These findings carry important implications for research and practice.
Our longitudinal evidence of how Cursor affects real-world software projects reveals complex temporal dynamics between AI-augmented velocity gains and quality outcomes (Figure~\ref{fig:theory}), warranting further investigation.
For practitioners, our results suggest that deliberate process adaptations---those that scale quality assurance with AI-era velocity---are necessary to realize sustained benefits from the use of LLM agent assistants.
Our findings also highlight the need for quality assurance as a first-class design citizen in AI-driven development tools and workflows, suggesting directions for improvement in tool design and model training.

\section{Related Work}

\label{sec:llm-impact}

The human-level performance of recent LLMs enables their practical applications to various software engineering tasks, such as code completion~\cite{DBLP:journals/csi/HuseinAC25}, code review~\cite{DBLP:conf/issre/LuYLYZ23}, and testing~\cite{DBLP:journals/tse/SchaferNET24} (see also the two surveys by~\citet{DBLP:conf/fose-ws/FanGHLSYZ23} and ~\citet{DBLP:journals/tosem/HouZLYWLLLGW24}).
The 2024 Stack Overflow Developer Survey shows that 76\% of all respondents are using or planning to use LLM tools in their development process~\cite{stackoverflow-survey-ai}.
This wide adoption raises two main questions for researchers: (1)~To what extent do LLMs improve developer productivity? (2)~To what extent should we trust the code generated by LLMs?

A large body of prior research on the productivity impact of LLMs focuses on \emph{code completion tools}---mostly the pre-agentic GitHub Copilot~\cite{DBLP:conf/pldi/0001KLRRSSA22, DBLP:conf/chi/Vaithilingam0G22, DBLP:conf/icse/Imai22, dohmke2023sea, DBLP:journals/corr/abs-2302-06590, hoffmann2024generative, DBLP:conf/icis/YeverechyahuMO24, DBLP:journals/corr/abs-2410-02091, cui2024effects, DBLP:journals/corr/abs-2406-17910, stray2025developer}, with only a few execeptions~\cite{DBLP:journals/corr/abs-2410-12944, kumar2025intuition}.
Evidence from small-scale, constrained randomized controlled experiments demonstrates a productivity increase ranging from 21\%~\cite{DBLP:journals/corr/abs-2410-02091}
to 56\%~\cite{DBLP:journals/corr/abs-2302-06590}, as measured by task completion time.
Field experiments conducted at Microsoft, Accenture, and Cisco report similar numbers (from 22\% \cite{cui2024effects} to 36\%~\cite{DBLP:journals/corr/abs-2406-17910}). 
The productivity increase estimated from open-source projects on observational data is similar and sometimes lower: a DiD design comparing Python and R packages estimates a 17.82\% increase in new releases among Python packages after Copilot availability~\cite{DBLP:conf/icis/YeverechyahuMO24}---without clear knowledge of which packages used Copilot.
Another study of proprietary Copilot backend data estimates only a 6.5\% increase in project-level productivity, as measured by the number of accepted pull requests~\cite{DBLP:journals/corr/abs-2410-02091}.
A more general study using a neural classifier to identify AI-generated code on GitHub shows moving to 30\% AI use raises quarterly commits by 2.4\%~\cite{daniotti2025using}.
Studies point to various mechanisms causing the productivity increase, such as how LLM adoption increases work autonomy~\cite{hoffmann2024generative} and helps iterative development tasks (e.g., bug fixing)~\cite{DBLP:conf/icis/YeverechyahuMO24}.

Although the productivity gains are promising, there are also increasing concerns around the trustworthiness of LLM-generated code.
For example, it is well-known that LLMs may generate code with security vulnerabilities~\cite{DBLP:conf/sp/PearceA0DK22, DBLP:conf/smc/KhouryABC23,  DBLP:journals/corr/abs-2310-02059, DBLP:conf/csr2/AmbatiRBS24, DBLP:journals/tse/LiuTLZZ24}, performance regressions~\cite{DBLP:conf/issre/LiC0X0S24}, code smells~\cite{DBLP:conf/scam/SiddiqMMJS22}, and outdated APIs~\cite{DBLP:journals/corr/abs-2502-01853, wang2025llms}.
On the other hand, evidence regarding the complexity of LLM-generated code compared to humans is inconclusive~\cite{DBLP:conf/msr/NguyenN22, DBLP:journals/corr/abs-2409-19182, martinovic2024impact}.
The LLM trustworthiness problem becomes more complicated with humans in the loop.
For example, prior controlled experiments report mixed results on whether developers write more or less secure code with the help of LLMs~\cite{DBLP:conf/ccs/PerryS0B23, DBLP:conf/sp/OhLPKK24, DBLP:conf/uss/SandovalPNKGD23}, and studies often suggest heterogeneous treatment effects of LLMs on developers of different skill levels~\cite{DBLP:journals/jss/DakhelMNKDJ23, DBLP:journals/corr/abs-2302-06590, DBLP:journals/corr/abs-2410-02091, cui2024effects}.
While prior work points to many mechanisms by which adopting LLMs may affect software quality, their findings are typically derived from benchmark analyses~\cite[e.g.,][]{DBLP:conf/sp/PearceA0DK22} or developer opinions~\cite[e.g.,][]{martinovic2024impact}.
We are unaware of any prior studies that systematically investigated project-level quality outcomes in the wild after LLM adoption, let alone those that used rigorous causal inference techniques (the closest being the study by \citet{DBLP:conf/icis/YeverechyahuMO24} discussed above).

Recently, there has been an increasing interest in the application of \emph{LLM agents}---LLMs with the capability to autonomously utilize external resources and tools---to software engineering~\cite{DBLP:journals/corr/abs-2408-02479, DBLP:journals/corr/abs-2404-04834, DBLP:journals/corr/abs-2409-02977}.
A popular application scenario is an \emph{LLM agent assistant} within a code editor, in which LLMs are allowed to inspect/edit project files, conduct web searches, and execute shell commands to fulfill prompts provided by developers.
At the time of writing, there are several production-ready code editors with built-in LLM agent assistants, such as Cursor~\cite{cursor}, VS Code~\cite{vscode}, Windsurf~\cite{windsurf}, Tabnine~\cite{tabnine}, and Cline~\cite{cline}, with the extreme beginning to shift away from IDEs entirely and switching to command-line or web interfaces, such as Claude Code~\cite{claude-code} and OpenHands~\cite{openhands}.
These agentic tools are seeing rapid adoption among developers, as evidenced also in our data for Cursor in Figure~\ref{fig:time-cursor-adoption}.
From the gray literature, we see extremely optimistic estimates of the productivity boost from LLM agent assistants: for example, developers self-report multi-fold productivity increases in a Reddit post~\cite{cursor-reddit-survey}, orders of magnitude larger than any empirical estimates for prior LLM tools.
However, a recent controlled study with human participants shows that developers may be overoptimistic and that adopting LLM agent assistants does not make them faster in real-world open-source development tasks~\cite{DBLP:journals/corr/abs-2507-09089}.
To the best of our knowledge, empirical evidence regarding the impact of LLM agent assistants on \emph{long-term project-level outcomes}, especially \emph{software quality outcomes}, is still lacking.

Our contribution is two-fold. 
First, our DiD design looks at the additional project-level productivity gain, if any, from using a modern \textit{agentic} coding assistant (Cursor) \emph{relative to the state-of-the-practice} (likely a mixture of human-written code and code generated by earlier-generation AI tools).
Second, we provide a comprehensive analysis of the impact of adopting Cursor on code quality, which is the first to the best of our knowledge, and highlight potential velocity-quality trade-offs and their complex interactions.

\section{Research Design and Methods}

We estimate the \emph{causal} effects of adopting Cursor on \emph{development velocity} and \emph{code quality}, both of which are considered important project outcomes, are commonly measured in prior research, and are closely tied to perceived overall project productivity~\cite{DBLP:journals/queue/ForsgrenSMZHB21, DBLP:journals/ese/OliveiraFSCCG20, DBLP:conf/sigsoft/ChengMCJGKZ022}.
We start by building a dataset with: (1) repositories adopting Cursor at different times and (2) comparable repositories that never adopted Cursor (Section~\ref{sec:data-collection}).
Then, we define our specific outcomes of interest and additional covariates (Section~\ref{sec:metrics}).
The estimation of Cursor's causal effect on these outcomes is enabled by a difference-in-differences (DiD) design with staggered adoption~\cite{borusyak2024revisiting} (Section~\ref{sec:modeling}): Under the assumption that similar repositories would, on average, evolve similarly in the absence of Cursor adoption (i.e., the \emph{parallel trend assumption}), later-adopters and never-adopters can effectively serve as a quasi-experimental comparison group for earlier-adopters while accounting for observable covariates and macro trends (e.g., open-source repositories overall getting more or less active over time).
Finally, since the results from DiD suggest \emph{interactions} between development velocity and software quality, as also indicated in prior work~\cite{DBLP:conf/sigsoft/ChengMCJGKZ022, besker2018technical}, we fit dynamic panel generalized method of moments (GMM) models~\cite{arellano1991some} to support our interpretation of the DiD results (Section~\ref{sec:interaction}).

\subsection{Data Collection}
\label{sec:data-collection}

\subsubsection{The Cursor IDE}
\label{sec:cursor}

Cursor~\cite{cursor} is an AI-powered IDE built as a VS Code fork with agentic capabilities integrated into the development workflow. 
Unlike earlier code completion tools, Cursor's agentic mode enables an autonomous, goal-directed AI workflow: The agent can navigate entire codebases, infer project architecture across multiple files, make multi-file edits, run terminal commands, execute tests, and iteratively debug code, with humans mainly serving as supervisors rather than traditional coders. 
Using this workflow, developers can rely entirely on AI for feature implementation, refactoring, test generation, documentation, and bug fixing within their native development environment. 
They can choose between frontier models from OpenAI, Anthropic, and Google, either via Cursor's built-in service or their own API keys.

We chose Cursor as our main study case for two reasons. 
First, our preliminary exploration found widespread and growing adoption compared to competing tools, providing sufficient statistical power for causal inference. 
Second, Cursor is among the earliest to allow optional configuration files (e.g., \Code{.cursorrules}) in the git repository to direct AI behavior~\cite{cursor-rules}, leading to an adoption event proxy timestamp in the version control history and a scalable identification strategy based on these files.
However, this identification strategy also has important limitations (Section~\ref{sec:limitations}), which we will address through robustness checks (Section~\ref{sec:robustness}).

\subsubsection{Identifying GitHub Projects Adopting Cursor}
\label{sec:find-cursor-repo}

We identify Cursor-adopting repositories and track adoption dates through configuration files in the git history. 
In the GitHub code search API~\cite{github-search-api}, we query for repositories with \Code{.cursorrules} files or \Code{.cursor} folders. 
Since the API limits results to 1,000 per query, we implement an adaptive partitioning algorithm based on file sizes: For each query with size interval $[a, b]$, we create two queries $[a, (a+b)/2)$ and $[(a+b)/2, b]$ until results fall below 1,000. This discovered 23,308 Cursor files across 3,306 non-fork repositories as of March 2025.

To filter non-software, educational, toy, and spam repositories~\cite{DBLP:journals/ese/KalliamvakouGBS16, DBLP:journals/corr/abs-2412-13459}, we follow prior work~\cite{DBLP:journals/corr/abs-2502-06662, DBLP:journals/tse/HeHZZ23, DBLP:conf/sigsoft/Soto-ValeroDB21} by requiring at least 10 stars at collection time---a threshold achieving 97\% precision in identifying engineered projects~\cite{DBLP:journals/ese/MunaiahKCN17}. This yields \numtreated repositories with adoption dates between January 2024 and March 2025 that are still available on GitHub at the time of data analysis (August 2025).

\begin{figure}
    \centering
    \includegraphics[width=0.95\linewidth]{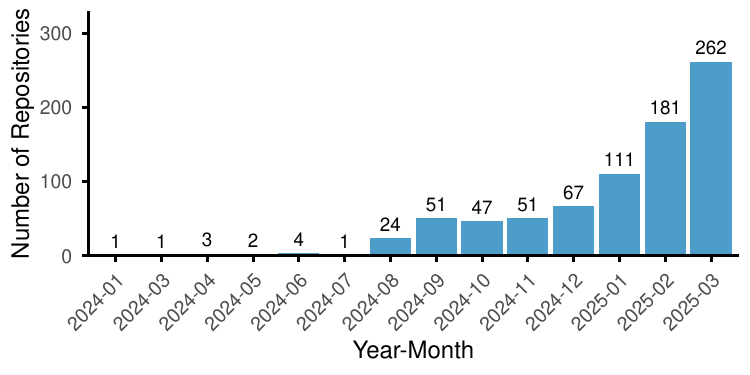}
    \caption{The Cursor adoption time of the \numtreated repositories in our study,  which all have $\ge$10 stars and Cursor configuration files at the time of data collection (April 2025).}
    \label{fig:time-cursor-adoption}
\end{figure}

As expected, the dataset is highly skewed across many repository-level metrics (Table~\ref{tab:repo-stats}), and adoption time is staggered, with adoptions growing over time (Figure~\ref{fig:time-cursor-adoption}).
These dataset characteristics motivate us to adopt a DiD design with staggered adoption and a matched control group, as we will discuss in the remainder of this section.
While we did not filter based on activity levels here, activity-based subsets will be used as part of our robustness checks.
The top five primary programming languages in our dataset are: TypeScript (366 repositories), Python (118 repositories), JavaScript (60 repositories), Go (36 repositories), and Rust (24 repositories).

\begin{table}
\small
\centering
\caption{Descriptive statistics of the \numtreated repositories using Cursor, collected at the time of data collection (April 2025).}
\label{tab:repo-stats}
\centering
\begin{threeparttable}
\begin{tabular}[t]{lrrrrrr}
\toprule
  & Mean & Min & 25\% & Median & 75\% & Max\\
\midrule
Age (days) & 1,002.3 & 283 & 361.5 & 521.5 & 1164.2 & 6,208\\
Stars & 1,475.3 & 10 & 20.0 & 51.0 & 242.0 & 122,280\\
Forks & 215.9 & 0 & 3.0 & 9.0 & 37.0 & 51,745\\
Contributors & 19.2 & 0\tnote{*} & 1.0 & 3.0 & 10.0 & 461\\
Commits & 1,816.6 & 1 & 49.0 & 209.0 & 951.8 & 86,954\\
Issues & 1,070.3 & 0 & 3.0 & 31.5 & 232.5 & 100,614\\
Pull Requests & 719.6 & 0 & 1.0 & 18.0 & 161.8 & 72,015\\
\bottomrule
\end{tabular}
\begin{tablenotes}
    \item[*]The GitHub API will return zero contributors for a repository if none of its commits can be mapped back to a GitHub user.
\end{tablenotes}
\end{threeparttable}
\end{table}

\subsubsection{Building a Control Group via Propensity Score Matching}
\label{sec:matching}
A staggered DiD design usually requires a ``never-treated'' group to serve as comparison units, which, in our case, means repositories that never adopted Cursor (i.e., \emph{never-adopters}).
This comparison allows for a causal interpretation under the parallel trend assumption: Repositories that adopted Cursor should, on average, evolve similarly in the absence of Cursor adoption compared to the control group.
However, this assumption is unlikely to hold if there are systematic differences between Cursor-adopting repositories and the control group.
Repositories adopting Cursor may be more active, more rapidly growing, and have larger communities compared to random never-adopter repositories, leading to biased estimates.

To address these potential confounders and build a comparable quasi-experimental control group, we use \emph{propensity score matching}~\cite{caliendo2008some, austin2011introduction}. 
The high-level idea is to estimate propensity scores (i.e., the probability of adoption conditional on pre-adoption covariates) for each Cursor-adopting repository and a large population of other GitHub repositories, and retain only never-adopter repositories with propensity scores similar to those of the control group. 
Specifically, we define this population as all GitHub repositories with $\geq$10 stars (matching our inclusion threshold for Cursor adopters). For each month with major Cursor adoptions (August 2024---March 2025), we collect monthly time series from GHArchive~\cite{gharchive} for all repositories in the population: age, active users, stars, forks, releases, pull requests, issues, comments, and total events.

Rather than static snapshots, we fit propensity score models to capture \textit{dynamics}: repositories experiencing rapid growth or changing patterns may be more likely to adopt new tools. Let $T_{it}$ denote repository age and $X_{it}$ denote remaining covariates for repository $i$ at month $t$.
We estimate propensity scores $P(\text{treat}|t, T_i, X_i)$ (i.e., the probability of getting treated given $t$ and repository-level covarates $T_i$, $X_i$) via the following logistic regression:
\begin{equation}
\log\frac{P(\text{treat}|...)}{1-P(\text{treat}|...)} = \alpha + \beta T_{i, t-1} + \sum_{j=1}^{6}\Gamma_j X_{i,t-j} + \Theta\sum_{j=7}^{\infty}X_{i,t-j} + \epsilon_{i}
\end{equation}
where $\alpha$, $\beta$, $\Gamma_j$ and $\Theta$ are regression parameters. This effectively captures: (1)~\textit{repository maturity} ($T_{i,t-1}$), as older repositories may have different adoption patterns, (2)~\textit{recent dynamics} ($\sum_{j=1}^{6}\Gamma_j X_{i,t-j}$), i.e., month-by-month evolution over six months captures trends and growth, and (3)~\textit{historical baselines} ($\Theta\sum_{j=7}^{\infty}X_{i,t-j}$), i.e., cumulative history providing context on overall project scale and activity.
By including lags, we ensure that propensity scores reflect both activity \textit{level} and \textit{trajectory}---two repositories with identical July 2024 pull requests may differ if one is growing while the other declines; such longitudinal characteristics may correlate with both Cursor adoption and long-term velocity and quality outcomes.

Since candidate repositories outnumber Cursor adopters by orders of magnitude, we sample at most 10,000 candidates per Cursor adoption month to avoid extreme imbalance and improve fit. This yields AUC values ranging from 0.83 to 0.91, indicating high discriminative power in the fitted logistic regression models.
For each Cursor-adopting repository, we perform 1:3 nearest-neighbor matching (three controls per treated unit). 
While 1:1 or 1:2 is most common~\cite{austin2010statistical, rassen2012one}, many adopters matched the same control during 1:1 matching, so 1:3 provides higher control group diversity. 
We additionally match only repositories with the same primary language (queried from GitHub API, unavailable in GHArchive), controlling for language-specific LLM performance differences~\cite{DBLP:journals/pacmpl/CassanoGLSFAF0J24, DBLP:journals/corr/abs-2503-17181}. 
This yields a matched ``never-treated'' group with similar propensity score distributions and pre-adoption characteristics (see Appendix).
Following standard quasi-experimental terminology, we refer to the sample of repositories adopting Cursor as the \emph{treatment group} and the matched ``never-treated'' repositories not using Cursor as the \emph{control group} in the remainder of this paper.

\subsection{Metrics}
\label{sec:metrics}

For each repository in the treatment and control group, we collect monthly outcome metrics and time-varying covariates from January 2024 to August 2025.
This ensures that there are at least six months of observations pre- and post-adoption for the treatment group and abundant comparison observations from the control group in each month, providing statistical power for DiD-based causal inference.
Note that the dataset is an \emph{unbalanced panel}, as not all repositories have observations over the entire observation period.

\subsubsection{Outcomes}
\label{sec:outcome-metrics}

For repository $i$ at month $t$, we collect two outcome metrics related to \textit{development velocity}, a key dimension of software engineering productivity~\cite{DBLP:journals/tse/Murphy-HillJSSP21, DBLP:conf/sigsoft/ChengMCJGKZ022}:
\begin{itemize}[leftmargin=15pt]
    \item Commits$_{it}$: Number of commits in repository $i$ at month $t$;
    \item Lines Added$_{it}$: Total lines added, summed over all commits in repository $i$ at month $t$.
\end{itemize}
Both have been used as productivity proxies~\cite{DBLP:journals/tosem/MockusFH02, DBLP:journals/jss/Lawrence81, DBLP:journals/ese/ScholtesMS16} with moderate-to-strong correlation with perceived productivity~\cite{DBLP:journals/ese/OliveiraFSCCG20}.

Software quality is multi-faceted and difficult to capture with a single metric~\cite{DBLP:journals/sigmetrics/CavanoM78, DBLP:journals/software/KitchenhamP96, DBLP:journals/csur/Osterweil96, fenton2014software}. Quality can be pivoted on defect density~\cite{DBLP:journals/cacm/Grady93}, specification rigor~\cite{farbey1990software}, user satisfaction~\cite{denning1992software}, or technical debt~\cite{DBLP:conf/sigsoft/ErnstBONG15}. 
However, many metrics cannot be reliably and scalably collected from version control data.
In this study, we take the technical debt perspective~\cite{martin2009clean} and test three source code maintainability metrics that can be reasonably estimated with static analysis: \emph{static analysis warnings}, \emph{duplicate line density}, and \emph{code complexity}. 
All three are arguably positively correlated with project-level technical debt and negatively correlated with perceived code quality. 
Specifically, we use a local SonarQube Community server~\cite{sonarqube} to compute these outcome metrics for repository $i$ at month $t$:
\begin{itemize}[leftmargin=15pt]
    \item Static Analysis Warnings$_{it}$: Total number of reliability, maintainability, and security issues for repository $i$ at month $t$, as detected by SonarQube's static analysis. We refer to them as warnings, since static analysis can generate false positives~\cite{DBLP:journals/tse/GuoTLLLYLCDZ23}; this metric should be viewed as an estimate of the effort required to review potential issues in a project.
    \item Duplicate Line Density$_{it}$: Percentage of duplicated lines in codebase for repository $i$ at month $t$. SonarQube's definition varies across programming languages, but usually requires at least 10 consecutive duplicate statements or 100 duplicate tokens to mark a block as duplicate~\cite{sonarqube-metrics}.
    \item Code Complexity$_{it}$: Overall cognitive complexity~\cite{DBLP:conf/icse/Campbell18} of codebase for repository $i$ at month $t$. Per SonarQube~\cite{DBLP:conf/icse/Campbell18}, this metric quantifies code understandability and aligns better with modern coding practices than classic cyclomatic complexity~\cite{DBLP:journals/tse/McCabe76}.
\end{itemize}

\subsubsection{Time-Varying Covariates}
\label{sec:covariates}

We control for the following time-varying covariates in our models for all treatment and control repositories over the entire observation period (Jan 2024 to Aug 2025): lines of code, age (days), number of contributors at month $t$, number of stars received at month $t$, number of issues opened at month $t$, and number of issue comments added at month $t$. Lines of code is collected from SonarQube~\cite{sonarqube} along with outcome metrics; number of contributors is estimated from version control history; remaining covariates are estimated from GHArchive event data~\cite{gharchive}. Multi-collinearity analysis reveals that number of issues opened and number of issue comments added are highly correlated (Pearson's $\rho > 0.7$), so we exclude issue comments from subsequent modeling.

\subsection{Difference-in-Differences}
\label{sec:modeling}

\subsubsection{Background}

DiD is an established econometric technique for causal inference in observational data~\cite{card1993minimum, cunningham2021causal}, with growing adoption in software engineering~\cite{nakasai2018donation, fang2022damn, casanueva2025impact}. 
The key idea is to compare outcome changes in a treatment group to those in a control group (i.e., those not-yet-treated or never-treated) over the same observation periods. 
The name ``difference-in-differences'' originates from the fact that temporal changes are first differenced before differencing the outcome changes between the two groups, effectively isolating the effect of an intervention from other factors that affect all repositories similarly over the same period.

A DiD design critically relies on the \textit{parallel trends assumption} for a causal interpretation: Absent treatment, the treatment and control group should, on average, follow similar outcome trajectories over the same period. 
While this assumption is generally not directly testable (would need a time machine), \emph{pre-trend tests} are often used for assessing the \emph{plausibility} of this assumption: If the model predicts similar outcomes for the treatment and control groups \emph{before the adoption}, it is more plausible that the treatment group would evolve similarly if they were not exposed to the treatment.
Apart from pre-trend tests, a matching process that controls for observable differences pre-adoption (e.g., Section~\ref{sec:matching}) also strengthens the plausibility of this assumption in a specific research context.

In a DiD design, a \textit{staggered adoption} setting comes with both promises and perils~\cite{athey2022design, borusyak2024revisiting}. 
In this setting, treatments occur at different times across cohorts rather than simultaneously (as in our case, Figure~\ref{fig:time-cursor-adoption}), and each treatment unit has repeated observations both before and after treatment.
This setting enables repeated, multiple natural experiments: Later-adopters in the treatment group can serve as additional controls for those adopting before them, since they remain untreated during earlier periods. 
However, a staggered adoption setting also presents significant mathematical challenges to achieve consistent, efficient, and unbiased estimation of the causal effect~\cite{de2020two, athey2022design, goodman2021difference} with ongoing active research~\cite{callaway2021difference, borusyak2024revisiting}.

\subsubsection{The Estimation Targets}
\label{sec:estimation-target}

In a DiD design, researchers are often interested in estimating the following two types of causal effects:
(1)~$ATT$, the average treatment effect on treated, and (2)~$ATT_h$, the ``horizon-average'' treatment effect in a specific horizon $h$ (i.e., the effect in $h$ periods since the treatment). 
We define the two estimation targets mathematically in this section.

Let $\Omega =\{it\}$ denote the set of all observations from repository $i$ and month $t$, $\Omega_1$ denote the set of treated observations, and $\Omega_0$ denote the set of untreated (i.e., never-treated and not-yet-treated) observations.
Let $Y_{it}$ denote the \emph{actual} outcome of interest for repository $i$ at month $t$, and $Y_{it}(0)$ denote the \emph{potential} outcome for repository $i$ at month $t$ if it is never treated. 
ATT is defined as the average of this causal treatment effect on all treated observations:
\begin{equation}
\label{eq:att}
ATT=\frac{1}{|\Omega_1|}\sum_{it\in\Omega_1}\mathbb{E}[Y_{it} - Y_{it}(0)]
\end{equation}
Let $\Omega_{1,h}$ denote the set of all treated observations $h$ time periods after the treatment; $ATT_h$ is defined as the average of the causal treatment effect in that specific horizon $h$: 
\begin{equation}
\label{eq:atth}
ATT_h=\frac{1}{|\Omega_{1,h}|}\sum_{it\in\Omega_{1,h}}\mathbb{E}[Y_{it} - Y_{it}(0)]
\end{equation}

\subsubsection{The \citet{borusyak2024revisiting} Estimator}
\label{sec:borusyak}

There are many possible methods to estimate $ATT$ and $ATT_h$ defined in Section~\ref{sec:estimation-target}, among which the two-way fixed effects (TWFE) estimator is most commonly used in early econometric studies.
However, recent research shows that TWFE may produce biased estimates in the staggered adoption setting if treatment effects are heterogeneous over time~\cite{de2020two, athey2022design, goodman2021difference}. 
To address this known limitation, we use the \emph{\citet{borusyak2024revisiting} imputation estimator}, a state-of-the-art estimator designed explicitly for robust and efficient estimation in the staggered adoption setting, with this two-step process:

\textit{Step 1: Impute counterfactual outcomes.} The estimator fits a counterfactual outcome regression model, \emph{using only untreated observations} $\Omega_0$ (i.e., pre-adoption observations for treated repositories and all observations for never-treated controls):
\begin{equation}
\label{eq:imputation}
\hat{Y}_{it}(0) = \hat{\mu}_i + \hat{\lambda}_t + \hat{\Gamma}' Z_{it} + \epsilon_{it}
\end{equation}
where $\hat{\mu}_i$, $\hat{\lambda}_t$ represent per-repository and per-month fixed effects; $Z_{it}$ includes time-varying covariates (Section~\ref{sec:covariates}), and $\epsilon_{it}=Y_{it}-\hat Y_{it}$, $it\in\Omega_0$ is the error term when fitting this model on $\Omega_0$. 
The use of only $\Omega_0$ in this step ensures that counterfactual predictions are not contaminated by treatment effects as in the TWFE estimator.
It is also important to note that all repository-invariant confounders (e.g., team culture, domain, language) are effectively controlled by per-repository fixed terms $\hat\mu_i$ and all time-invariant confounders (e.g., industry trends, platform changes, and seasonal patterns) are effectively controlled by per-month fixed terms $\hat{\lambda}_t$ in this step.

\textit{Step 2: Compare actual to counterfactual.} For each repository-month post-adoption ($it\in\Omega_1$), the estimator predicts the potential outcome from the counterfactual outcome model: $\hat{Y}_{it}(0) = \hat{\mu}_i + \hat{\lambda}_t + \hat{\Gamma}' X_{it}$. We replace the $Y_{it}(0)$  in Equations~\ref{eq:att} and~\ref{eq:atth} with the estimated $\hat{Y}_{it}(0)$, to get the final $ATT$ and $ATT_h$ estimations.

Finally, to assess the plausibility of the parallel trend assumption, \citet{borusyak2024revisiting} advises fitting an alternative model of $Y_{it}$ for untreated observations $\Omega_0$ with additional observables.
In our paper, we follow the most typical convention and fit the following model with additional dummies before the onset of treatment:
\begin{equation}
\label{eq:parallel-trend}
    Y_{it}=\hat{\mu}_i + \hat{\lambda}_t + \hat{\Gamma}' Z_{it} + \sum_{h=-k}^{-2}\hat\tau_h\mathbf{1}[t=E_i + h] + \epsilon_{it}
\end{equation}
where $E_i$ means the time of treatment for repository $i$ and $\mathbf{1}[t=E_i+h]$ is equal to $1$ if and only if the current time $t$ is $h$ months away from treatment (0 otherwise).
Then, we use heteroscedasticity- and cluster-robust Wald tests~\cite{borusyak2024revisiting} to test the joint null hypothesis that $\hat\tau_h=0$ for $h=-k,...,-2$.
We drop $h=-1$ due to potential anticipation concerns (the developer may try use Cursor before adding Cursor rule files in the immediate month before).
One way of viewing the above pre-trend testing procedure is a placebo test, in which $\tau_h$ estimates $ATT_h$ for $h < 0$, which should be zero if the treatment has not yet happened.
The estimated $\hat\tau_h$ for $h<0$ and $ATT_h$ for $h>0$ are often combined into \emph{event study plots} (e.g., Figure~\ref{fig:dynamic-te}), in which the dynamic effect of treatment is visualized.

Note that the \citet{borusyak2024revisiting} imputation estimator has many alternative specifications with varying assumptions, and we merely describe the version we used in our paper.
We refer interested readers to the original paper~\cite{borusyak2024revisiting} regarding alternative specifications and the mathematical assumptions behind them.
The \citet{borusyak2024revisiting} estimator is also not the only option, and we present results from alternative DiD estimators (TWFE and \citet{callaway2021difference}) in the Appendix as additional robustness checks.


\subsection{Testing Velocity \& Quality Interactions}
\label{sec:interaction}

While DiD estimates treatment effects on individual outcomes, it does not capture temporal dynamics \textit{between} outcomes. 
However, it is known that velocity and quality outcomes interact in our setting~\cite{DBLP:conf/sigsoft/ChengMCJGKZ022, besker2018technical}. 
Plus, our DiD results (Section~\ref{sec:results}) also suggest interactions, showing that the adoption of Cursor leads to non-sustained velocity increases and sustained quality declines: Development velocity increases may cause rapid technical debt accumulation, which may subsequently decrease velocity.

To test for dynamic relationships and bidirectional causality, we use the \emph{generalized method of moments} (GMM)~\cite{hansen1982large} to obtain consistent estimates when variables are potentially endogenous (correlated with unobserved errors). 
The key insight is to use \emph{instrumental variables}---correlated with endogenous regressors but uncorrelated with errors---to identify causal effects.
In panel data, lagged values serve as natural instruments, assuming past values influence current values but are uncorrelated with current shocks~\cite{arellano1991some}.

In our study, we use Arellano-Bond dynamic panel GMM~\cite{arellano1991some}, suited for: (1)~dynamic dependence (current outcomes depend on past); (2)~potential bidirectional causality; (3)~short time series with many entities. 
To test a causality direction $X_t \to Y_t$ while accounting for Cursor adoption $D$, we estimate the following regression:
\begin{equation}
\label{eq:dynamic-gmm}
Y_{it} = \hat\mu_i + \hat\lambda_t + \hat\rho Y_{i,t-1} + \hat\beta D_{it} + \hat\gamma X_{it} + \hat{\Gamma}' Z_{it} +\epsilon_{it}
\end{equation}
where $Y_{i,t-1}$ captures outcome persistence; $D_{it}$ is a dummy representing Cursor adoption; $X_{it}$ represents the potentially endogenous regressor of interest; the remainig follows Equation~\ref{eq:imputation}. During estimation, historical values of $X_{it}$ (e.g., a linear combination of $X_{i,t-2}$ and $X_{i,t-3}$) are used as instrumental variables.

Specifically, we test the following temporal interactions:
\begin{equation}
\label{eq:temporal}
\begin{aligned}
\text{Lines Added}_{it} &\to \text{Static Analysis Warnings}_{it}\\
\text{Lines Added}_{it} &\to \text{Code Complexity}_{it}\\
\text{Static Analysis Warnings}_{it} &\to \text{Lines Added}_{i,t+1}\\
\text{Code Complexity}_{it} &\to \text{Lines Added}_{i,t+1}
\end{aligned}
\end{equation}
These models complement DiD by decomposing the mechanisms through which Cursor affects long-term outcomes, revealing whether quality degradation leads to subsequent velocity declines.

\subsection{Limitations and Threats to Validity}
\label{sec:limitations}

\subsubsection{Internal Validity}
We discuss several important limitations of our identification strategy and advise readers to interpret our results in the context of this experimental setup and its limitations. 

\emph{Observable adoption through committed configuration files.} Our treatment group only includes repositories committing Cursor configuration files to their git system, but developers can use Cursor without committing such files.
Thus, our sample represents repositories with observable Cursor adoption rather than all possible Cursor-adopting repositories. 
This creates potential selection bias: Repositories committing configuration files may be more committed to systematic integration, have more formal processes, or differ in unobservable ways. 
To the extent committed adopters use Cursor more systematically, our estimates may represent an upper bound on average effects across all users. However, if committed adopters are more quality-conscious (e.g., more likely to review AI-generated code carefully), estimates could also be conservative. 
In general, we view our sample as capturing repositories where adoption represents deliberate, visible practice change---precisely the population where long-term effects are most relevant. 

\looseness=-1
\emph{Uncertainty about usage intensity and persistence.} 
Even with observed configuration files, we do not know how intensively or persistently Cursor was used in each repository. 
If a repository has multiple contributors with their own development environment, the presence of \Code{.cursorrules} only indicates \emph{someone} experimented, not that \emph{all} contributors used it continuously throughout post-adoption observations. 
Contributors may use Cursor heavily for one feature, then revert to traditional development, without leaving visible traces. Unless we observe explicit configuration removal (rare), we assume continued usage, but this approximates actual engagement. 
Therefore, our main estimates represent intent-to-treat (ITT) effects: the impact of adopting Cursor as measured by committing configuration, averaging over heterogeneous usage patterns.

\emph{Model and version heterogeneity.} 
Our dataset lacks information on which Cursor version or LLM backend each repository used; regardless, we still argue that this does not compromise validity. 
Our research question focuses on the system-level effects of adopting an agentic coding assistant as an integrated development practice, rather than the effects of particular model architectures. 
Moreover, to the extent that different repositories use different model backends or developers switch models for different tasks, this heterogeneity increases external validity: Our estimates reflect the average treatment effects of adopting Cursor as it is actually used in practice on adopting repositories, across all model diversity, rather than the effects of single, fixed LLM configurations.

\emph{Imperfect matching.} 
Even if our propensity score matching process achieved strong performance (AUC 0.83--0.91), it still remains subject to untestable unobserved confounders. 
For example, factors like developer expertise, team practices, project complexity, or organizational culture may affect both Cursor adoption and the post-adoption outcomes. 
Although we include numerous covariates hoping to control these factors latently, we believe perfect matching is generally impossible in our setting~\cite{austin2011introduction}.
The matching process is intended to strengthen the plausibility of our DiD design, particularly the parallel trend assumption, rather than creating two directly comparable groups.
The DiD design further addresses imperfect matching through incorporating per-repository and per-period fixed effects (Equation~\ref{eq:imputation}) and time-varying covariates (Section~\ref{sec:covariates}) into the counterfactual outcome model.

\emph{Contamination from alternative AI coding tools.} 
Another particular concern for our study setting is that repositories in both the treatment and control groups may be using alternative AI tools before and throughout the observation period, with or without visible traces.
For example, many developers may use early versions of GitHub Copilot or chat-based interfaces like the ChatGPT web portal~\cite{stackoverflow-survey-ai} without leaving any visible traces in the git repository.
Therefore, our estimates should be interpreted as the impact of systematic Cursor adoption compared to the current state-of-the-practice, in which code completion tools and chat-based AI interfaces may be prevalently used, not the impact of using Cursor with respect to no AI usage at all (the latter is generally not estimable in our observational dataset).
However, we identify several observable AI coding tools in our dataset (e.g., Claude Code with \Code{.claude} folders) and present robustness checks in Section~\ref{sec:robustness}.

\subsubsection{External Validity}

Our results may not generalize to other LLM agent assistants, proprietary software projects, and programming languages beyond the three dominant ones in our dataset (JavaScript, TypeScript, Python)---adoption patterns and impacts may differ substantially in these contexts. Importantly, our study period coincides with rapid evolution in LLM capabilities, agent tooling, and developer adoption patterns. Results observed may not persist as LLM agent assistants mature and developer workflows adapt. We encourage future replications and additional investigations of state-of-the-art LLM coding tools as they roll out.

\section{Results}
\label{sec:results}

\subsection{Difference-in-Differences}

\begin{table}[t]
\centering
\small
\caption{The \citet{borusyak2024revisiting} estimated average treatment effects on treated ($ATT$ in Equation~\ref{eq:att}) post Cursor adoption. 
We log-transform all outcome variables to address skewness and facilitate comparison of treatment effects across outcome variables; after log-transformation, all estimated $ATT$s can be interpreted as a percentage change of $100(e^{ATT}-1)\%$.}
\begin{tabular}{lrr}
\toprule
Outcome & Estimate (Std. Error) & Percentage Change\\
\midrule
Commits & 0.0260$^{\textcolor{white}{***}}$ (0.0429) & +2.63\% ($\pm$4.40\%)\\
Lines Added & 0.2514$^{*\textcolor{white}{**}}$ (0.1063) & +28.58\% ($\pm$13.7\%)\\
Static Analysis Warnings & 0.2644$^{***}$ (0.0511) & +30.26\% ($\pm$6.66\%)\\
Duplicated Lines Density & 0.0679$^{\textcolor{white}{***}}$ (0.0448) & +7.03\% ($\pm$4.79\%)\\
Code Complexity & 0.3481$^{***}$ (0.0538) & +41.64\% ($\pm$7.62\%)\\
\bottomrule
\multicolumn{3}{r}{\textit{Note:} $^{*}p<0.05$; $^{**}p<0.01$; $^{***}p<0.001$} \\
\end{tabular}
\label{tab:average-te}
\end{table}

\begin{figure*}
    \centering
    \includegraphics[width=0.97\linewidth]{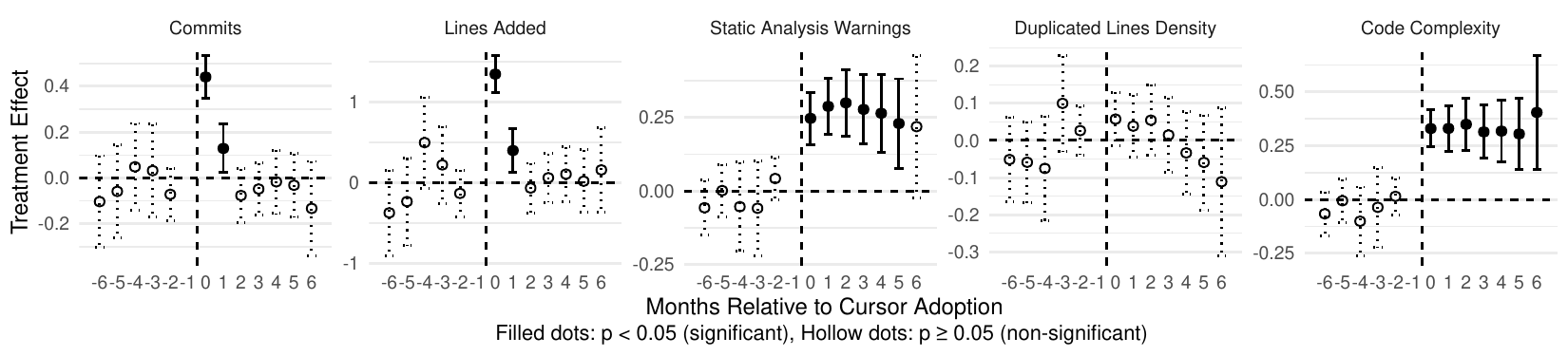}
    \caption{The estimated ``horizon-average'' treatment effects ($ATT_h$, Equation~\ref{eq:atth}) 0 to +6 months after adoption, plus the ``placebo'' pre-adoption treatment effect estimates ($\hat\tau_h$ for testing the parallel trend assumption, Equation~\ref{eq:parallel-trend}) -6 to -2 months before adoption. All outcome variables are log-transformed same as the estimated average treatment effects in Table~\ref{tab:average-te}.}
    \label{fig:dynamic-te}
\end{figure*}

We summarize the \citet{borusyak2024revisiting} estimated average treatment effects $ATT$ and horizon-average treatment effects $ATT_h$ (see definitions in Section~\ref{sec:modeling}) in Table~\ref{tab:average-te} and Figure~\ref{fig:dynamic-te} for the five velocity and quality outcome variables (see definitions in Section~\ref{sec:outcome-metrics}).
All outcome variables pass the heteroscedasticity- and cluster-robust Wald tests~\cite{borusyak2024revisiting} at the 0.05 level (i.e., passing pre-trend tests).

\subsubsection{Development Velocity}
\label{sec:finding-velocity}

On average, Cursor adoption has a modestly significant positive impact on development 
velocity, particularly in terms of code production volume: Lines added increase by about 28.6\% (Table~\ref{tab:average-te}).
There is no statistically significant effect for the volume of commits. 
The horizon-average treatment effect estimations (Figure~\ref{fig:dynamic-te}) reveal 
important temporal patterns that explain these differences: \emph{The only significant development velocity gain is in the first two months post Cursor adoption.}
The models estimate a 55.4\% increase in commits in the first month, a 14.5\% increase in commits in the second month, a 281.3\% increase in lines added in the first month, and a 48.4\% increase in lines added in the second month, respectively. 


\subsubsection{Software Quality}
\label{sec:finding-quality}

In contrast to the transient velocity gains, Cursor adopters show 
sustained patterns across static analysis warnings and code complexity.
On average (Table~\ref{tab:average-te}), static analysis warnings increase significantly by 30.3\%, and code complexity increases by 41.6\%.
The effect on duplicate line density is insignificant.
The horizon-average treatment effect estimates (Figure~\ref{fig:dynamic-te}) reveal that the increase in the two outcome metrics persists beyond the initial adoption period, contrasting the transient velocity gains. 


\subsection{Velocity \& Quality Interactions}
\label{sec:finding-interaction}

\begin{table}
\centering
\caption{The dynamic panel GMM estimates testing temporal interactions between velocity and quality attributes. $L$, $W$, and $C$ stand for lines added, static analysis warnings, and code complexity, respectively (see Equation~\ref{eq:temporal}). The estimates for the remaining covariates are omitted here for brevity.}
\label{tab:causal-paths}
\begin{threeparttable}
\small
\begin{tabular}{lrrrr}
\toprule
& \makecell{$L_{it} \to$$W_{it}$} & \makecell{$L_{it} \to$$C_{it}$} & \makecell{$C_{it} \to$$L_{i,t+1}$} & \makecell{$W_{it} \to$$L_{i,t+1}$} \\
\midrule
Main Effect & $-$0.000$^{\textcolor{white}{***}}$ & $-$0.006$^{\textcolor{white}{***}}$ & $-$0.718$^{***}$ & $-$0.588$^{***}$ \\
& (0.015)$^{\textcolor{white}{***}}$ & (0.016)$^{\textcolor{white}{***}}$ & (0.098)$^{\textcolor{white}{***}}$ & (0.092)$^{\textcolor{white}{***}}$ \\
[0.1em]
Cursor & $-$0.011$^{\textcolor{white}{***}}$ & 0.086$^{**\textcolor{white}{*}}$ & 1.044$^{***}$ & 1.048$^{***}$ \\
& (0.033)$^{\textcolor{white}{***}}$ & (0.030)$^{\textcolor{white}{***}}$ & (0.124)$^{\textcolor{white}{***}}$ & (0.124)$^{\textcolor{white}{***}}$ \\
[0.1em]
Lines of Code & 0.845$^{***}$ & 0.852$^{***}$ & 0.869$^{***}$  & 0.851$^{***}$  \\
& (0.073)$^{\textcolor{white}{***}}$ & (0.059)$^{\textcolor{white}{***}}$ & (0.153)$^{\textcolor{white}{***}}$	& (0.155)$^{\textcolor{white}{***}}$ \\
\midrule
Num. Obs. & 14,755$^{\textcolor{white}{***}}$ & 14,755$^{\textcolor{white}{***}}$ & 14,755$^{\textcolor{white}{***}}$ & 14,755$^{\textcolor{white}{***}}$ \\
Sargan $p$ & 0.248$^{\textcolor{white}{***}}$ & 0.141$^{\textcolor{white}{***}}$ & 0.633$^{\textcolor{white}{***}}$ & 0.639$^{\textcolor{white}{***}}$ \\
AR(1) $p$ & <0.001$^{\textcolor{white}{***}}$ & <0.001$^{\textcolor{white}{***}}$ & <0.001$^{\textcolor{white}{***}}$ & <0.001$^{\textcolor{white}{***}}$ \\
AR(2) $p$ & 0.734$^{\textcolor{white}{***}}$ & 0.438$^{\textcolor{white}{***}}$ & 0.393$^{\textcolor{white}{***}}$ & 0.330$^{\textcolor{white}{***}}$ \\
\bottomrule
\end{tabular}
\begin{tablenotes}
\small
\item \textit{Notes:} Two-way fixed effects (repository + month), two-step GMM with first-difference transformation. 
Robust standard errors in parentheses. ***$p$<0.001, **$p$<0.01, *$p$<0.05. The contemporaneous variables $L_{it}$, $C_{it}$, and $W_{it}$ are instrumented with lags 2-3 to address endogeneity. Sargan $p$>0.05 indicates valid instruments. AR(1) $p$<0.05 is expected with the first-difference transformation. AR(2) $p$>0.05 indicates no serial correlation in the original errors. Sargan $p$>0.05 and AR(2) $p$>0.05 validate the moment conditions required for causal interpretation~\cite{arellano1991some}.
\end{tablenotes}
\end{threeparttable}
\end{table}

To distangle the temporal interactions between velocity and quality, we summarize dynamic panel GMM models testing causal paths specified in Equation~\ref{eq:temporal}, in Table~\ref{tab:causal-paths}.
All models pass the Sargan test (confirming instrument validity) and AR(2) test (confirming no serial correlation in the original errors), validating the moment conditions required for causal interpretation~\cite{arellano1991some}.

The first two models show that, on average (across all Cursor-adopters and non-adopters in our dataset), and \emph{holding all other temporal dynamic factors constant}: 
(1) An increase in development velocity does not produce a significant effect on static analysis warnings and code complexity.
(2) Cursor adoption does not have a significant effect on static analysis warnings.
Notably, increases in codebase size are a major determinant of increases in static analysis warnings and code complexity, and absorb most variance in the two outcome variables.
However, even with strong controls for codebase size dynamics, the adoption of Cursor still has a significant effect on code complexity, leading to a 9\% baseline increase on average compared to projects in similar dynamics but not using Cursor.

The last two models show that, on average, and \emph{holding all other temporal dynamic factors constant}: (1) A 100\% increase in code complexity and static analysis warnings causes a 64.5\% and 50.3\% decrease in development velocity as measured by lines added, respectively.
(2) The adoption of Cursor results in a 1.84x baseline increase in lines added post adoption.
Thus, the velocity gain from Cursor adoption would be fully cancelled out by a $\sim$5x increase in static analysis warnings or a $\sim$3x increase in code complexity, according to the dynamic panel GMM estimations.

\begin{figure*}
    \centering
    \includegraphics[width=\linewidth]{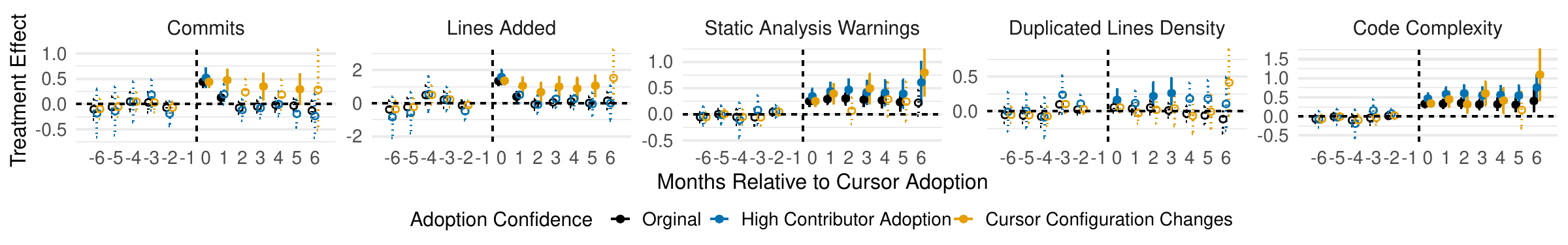}
    \includegraphics[width=\linewidth]{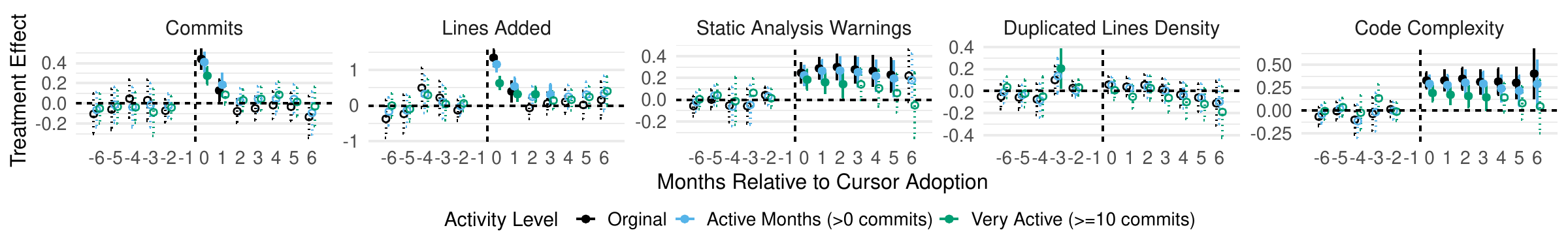}
    \includegraphics[width=\linewidth]{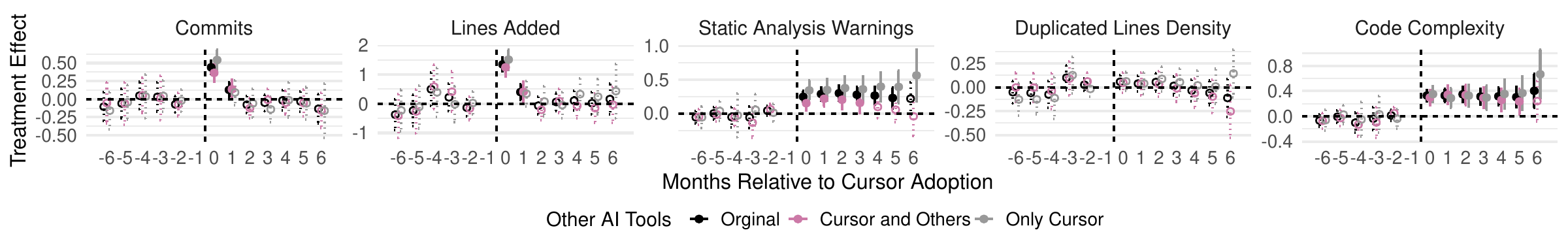}
    \caption{The estimated ``horizon-average'' treatment effects in alternative dataset settings as robustness checks. 
    \emph{Row 1}: Robustness check on subsets with higher Cursor adoption confidence, showing generally stronger effects in high-confidence adoption observations. 
    \emph{Row 2}: Robustness check on subsets of different activity levels, showing the persistence of our findings even in very active repositories. 
    \emph{Row 3}: Robustness check between repositories with/without observable use of other AI tools, showing that the potential use of other AI tools weakens Cursor effect estimations, but only by a small margin.}
    \label{fig:robustness-checks}
\end{figure*}

To summarize, the dynamic panel GMM models suggest that: (1) \emph{The adoption of Cursor leads to an inherently more complex codebase}; and (2) \emph{The accumulation of static analysis warnings and code complexity decreases development velocity in the future}.


\subsection{Robustness Checks}
\label{sec:robustness}

The above overall findings and the internal validity concerns behind them (Section~\ref{sec:limitations}) present significant interpretation challenges.
In this section, we present additional robustness checks to explore and rule out alternative explanations for our observed results.

Recall that one limitation of our identification strategy is that it merely displays an ``intent-to-treat'' (ITT) signal without us knowing how Cursor was actually used in each treatment repository (Section~\ref{sec:limitations}).
To test whether our findings are driven by repositories with genuine, sustained usage (versus minimal engagement), we repeat the same DiD modeling process on two subsets of data based on two alternative measurements of usage intensity:

\begin{itemize}[leftmargin=15pt]
    \item \emph{High Contributor Adoption:} For each repository, we identify contributors modifying Cursor files (indicating experimentation) and calculate their fraction of total repository commits during observation. 
    This subset keeps only ``high-adoption'' repositories, in which those modifying Cursor files also contributed $\geq$80\% of commits, along with their own matched controls.
    \item \emph{Cursor Configuration Changes:} For each repository, we identify all commits where \Code{.cursorrules} files were modified post-adoption. 
    This subset keeps only post-adoption observations with at least one commit modifying Cursor files during that period, representing sustained usage and ruling out potential post-adoption abandonments.
\end{itemize}

Results from the two subsets (Figure~\ref{fig:robustness-checks}, Row 1) show that our quality-related findings are robust and amplified in subsamples with higher sustained usage: The accumulation of static analysis warnings and code complexity is stronger---not weaker---among repositories with continued configuration refinement and where Cursor users dominated activity. 
This strengthens causal interpretation: Effects are attributable to Cursor adoption rather than spurious correlation with other changes, and our main ITT estimates likely understate effects among repositories with intensive, sustained usage.
Interestingly, while the velocity gain remains transient on average across repositories with high contributor adoption, we still observe a velocity increase in observations where developers are actively modifying Cursor files.
This indicates that abandonments of Cursor post-adoption likely nullify at least part of the velocity gains (more discussions in Section~\ref{sec:discussion}).

Another alternative explanation for the transient velocity gain is that our DiD setting still does not sufficiently control for the macro-trend that many open-source repositories become inactive very fast (e.g., within two months).
To test against this alternative explanation, we build two subsets of observations: The first keeps only repository-months with at least one commit, and the second keeps only repository-months with at least 10 commits.
The results (Figure~\ref{fig:robustness-checks}, Row 2) show that our main findings persist even in very active repositories, albeit weaker than our main ITT estimates, which is expected because it would be more difficult to achieve the same percentage change in the measured outcomes in repositories with higher baseline sizes and activity levels.

Finally, we assess the impact of dataset contamination from alternative AI coding tools in our study setting (Section~\ref{sec:limitations}).
For this purpose, we take an overly conservative approach and identify all repositories where alternative AI coding tools may have been used based on repository files (e.g., those with \Code{.vscode} folder may have used GitHub Copilot).
We find 382 repositories from the treatment group that may use other AI tools during our observation period: 345 with GitHub Copilot, 63 with Claude Code, 37 with Windsurf, 13 with Cline (13), and 2 with OpenHands (note that they overlap heavily).
We build a \emph{Cursor and Other} subset for these repositories and an \emph{Only Cursor} subset for the remaining repositories, along with their own matched controls.
The results (Figure~\ref{fig:robustness-checks}, Row 3) show that while contamination weakens our main ITT estimates (e.g., some repositories may be using GitHub Copilot before Cursor), all of our main findings are persistent through all three settings and amplified in settings where prior AI tool usage is less likely. 

To summarize, the above robustness checks reassure our main causal findings against concerns from potential non-compliance (i.e., treatment repositories not actually using Cursor), selection bias (i.e., treatment repositories generally becoming inactive fast), and confounding from other AI tools (i.e., the presence of other AI tools nullifying estimates on the adoption of Cursor).

\section{Discussion}
\label{sec:discussion}

\subsection{Theoretical Implications}

Our study contributes to the rapidly growing literature regarding the impact of AI assistance on developer productivity~\cite{DBLP:conf/pldi/0001KLRRSSA22, DBLP:conf/chi/Vaithilingam0G22, DBLP:conf/icse/Imai22, dohmke2023sea, DBLP:journals/corr/abs-2302-06590, hoffmann2024generative, DBLP:conf/icis/YeverechyahuMO24, DBLP:journals/corr/abs-2410-02091, cui2024effects, DBLP:journals/corr/abs-2406-17910, kumar2025intuition,
stray2025developer, DBLP:journals/corr/abs-2507-09089}.
More importantly, our study provides a novel longitudinal lens into project-level macro-outcomes, effectively connecting our findings to the existing software engineering literature around development velocity and software quality~\cite{DBLP:conf/sigsoft/ChengMCJGKZ022, storey2022developers, DBLP:journals/tse/Murphy-HillJSSP21, besker2018technical, kruchten2012technical}.
In this section, we connect our findings with prior research and discuss our theory around how LLM agent assistants may impact software development (Figure~\ref{fig:theory}).

\subsubsection{The Transient Velocity Gains and Possible Causes Behind It}

Our first longitudinal finding---that the project-level velocity gains from adopting Cursor are concentrated in the initial one or two months before returning to a baseline level---contrasts task-level productivity improvements reported in controlled experiments~\cite[e.g.,][]{DBLP:journals/corr/abs-2302-06590, DBLP:journals/corr/abs-2410-12944}. 
One reason for such contrast likely stems from the temporal dynamics between development velocity and software quality (which is only observable in a longitudinal study setting): While LLM agent assistants increase development velocity, the increase in development velocity itself may increase codebase size and cause accumulation of technical debt; the latter would consequently decrease development velocity in the future.
This negative effect of technical debt is supported by both the prior literature~\cite{DBLP:conf/sigsoft/ChengMCJGKZ022, besker2018technical} and our panel GMM models (Table~\ref{tab:causal-paths}).
However, this mechanism alone likely does not fully explain why the development velocity gain vanishes after two months: A $\sim$3x increase in code complexity or a $\sim$5x increase in static analysis warnings would be necessary to fully cancel out the effect of Cursor adoption according to our models (Table~\ref{tab:causal-paths}), which is unlikely.

Another highly plausible explanation, as indicated in Section~\ref{sec:robustness}, is that open-source developers experience an \emph{excitement-frustration-abandonment cycle} while they adopt Cursor.
For example, during the initial adoption phase, developers may experience novelty effects and actively experiment on tasks where AI excels (e.g., rapid prototyping), contributing to the immediate velocity spike post-adoption. 
However, as developers encounter scenarios where AI is still limited (e.g., debugging intricate logic, understanding existing codebases, handling edge cases), frustration may accumulate. 
This frustration, combined with the cognitive overhead of verifying and debugging AI-generated suggestions, could lead to reduced usage or complete abandonment.
This interpretation aligns with the emerging qualitative research documenting 
developer challenges with AI-assisted coding~\cite{DBLP:conf/icse/Liang0M24, DBLP:journals/corr/abs-2507-09089, DBLP:journals/corr/abs-2507-08149} and anecdotal evidence from Cursor users~\cite[e.g.,][]{cursor-critique1, cursor-critique2}.

\subsubsection{The Accumulation of Technical Debt and Code Complexity}

Our findings reveal a nuanced relationship between velocity and quality that 
challenges simplistic narratives about AI coding degrading code quality~\cite{gitclear-report}. 
While the absolute levels of static analysis warnings increase post adoption (Finding 2), a large part of this observed effect can be attributed to the causal path of increased velocity $\to$ increased code base size $\to$ increased technical debt (Table~\ref{tab:causal-paths}). 
In other words, LLM agent assistants amplify existing velocity-quality dynamics by enabling faster code production, but may not necessarily introduce more code quality issues than non-adopting projects moving with the same velocity.
This proportional relationship has an important practical implication that quality assurance needs to scale with AI-era velocity (see Section~\ref{sec:practical-implications}).
After all, the use of AI does not change the fact that all code is a liability and the asset lies only in the code's capabilities~\cite{code-liability-argument}.

The substantial average increase in code complexity (25.1\%, Table~\ref{tab:average-te}) warrants particular attention, as code complexity represents a distinct quality dimension from code quality issues. 
That code complexity increases even after accounting for velocity dynamics (Table~\ref{tab:causal-paths}) gives strong evidence that code generated with Cursor in our study sample is inherently more complex than human-written code.
This effectively creates a ``complexity debt'' in projects that use AI heavily, which may amplify frustration and maintenance costs when the AI fails on more complex codebases later, possibly providing another mechanism explaining the transient velocity gain we observe after Cursor adoption in our dataset.

While the adoption of Cursor leads to no significant changes in duplicate line density in the entire study sample (Table~\ref{tab:average-te}, Figure~\ref{fig:dynamic-te}), heavy Cursor adopters may exhibit modest increases (Figure~\ref{fig:robustness-checks}).
Future research is needed to gather evidence on code duplication concerns in high AI usage scenarios.

\subsubsection{Contextual Factors in Open-Source Settings}

Our findings should be interpreted within the specific context of open-source software development, which differs from enterprise settings in ways that likely influence the patterns we observe. 
Open-source projects typically feature: (1) voluntary participation with low switching costs, enabling easy abandonment when tools prove frustrating, (2) distributed collaboration with varying levels of coordination, potentially reducing systematic code review that might catch AI-introduced defects; (3) intrinsic motivation and learning goals, where experimenting with AI tools provides value beyond pure productivity; and (4) resource constraints that may limit comprehensive testing and quality assurance regardless of development velocity.
These contextual factors likely amplify the excitement-frustration-abandonment cycle while potentially dampening the quality feedback loop. 
In enterprise settings, organizational mandates, sunk training costs, and managerial oversight might sustain AI tool use despite user frustration, potentially leading to distinct temporal patterns (as indicated in a recent study~\cite{kumar2025intuition}). 
Similarly, enterprise quality assurance processes---mandatory code review, automated testing requirements, dedicated QA teams, and even dedicated agents to do maintenance work---might prevent proportional technical debt accumulation by catching issues before they escalate. 
Future research should examine whether the transient gains and proportional debt patterns we observe generalize to enterprise contexts or represent open-source-specific phenomena.

\subsection{Practical Implications}
\label{sec:practical-implications}

\emph{To overcome the technical debt accumulation ratchet, software projects using LLM agent assistants should focus on process adaptation that scales quality assurance with AI-era velocity}.
The proportional technical debt accumulation we observe (Section~\ref{sec:finding-quality}), combined with its velocity-dampening effects (Section~\ref{sec:finding-interaction}), creates a self-reinforcing cycle that needs to be addressed at the project level. 
To overcome this, AI-adopting teams may consider refactoring sprints triggered by code quality metrics, mandating test coverage requirements that scale with lines of code added, or prompt engineering (e.g., engineered Cursor rules) to enforce rigid quality standards for LLM agents. 
Without such adaptations, the initial productivity surge may accelerate the project toward an unmaintainable end state. 

\emph{To support the above process adaptation, AI coding tools need explicit design to support quality assurance alongside code generation.}
The LLM agents (at the time of study) are generation-first, leaving quality maintenance as an afterthought.
Future assistants should suggest tests alongside code, flag unnecessary complexity in real time, and proactively recommend refactoring when code quality degrades---essentially becoming ``pair programmers'' for quality, not just velocity. 
More provocatively, tools might implement self-throttling: automatically reducing suggestion volume or aggressiveness when project-level complexity or debt exceeds healthy thresholds, forcing developers to consolidate before generating more code. 
Such features would align tools with long-term project health rather than short-term code production.

\emph{The potential overcomplication in AI-generated code warrants further research and improvement.}
The 25\% increase in code complexity we observe (Table 3) represents a distinct quality dimension beyond code quality issues---a ``comprehension tax'' that persists regardless of functional correctness. 
This suggests LLMs may be generating structurally valid but semantically opaque code, perhaps because training objectives prioritize passing tests over non-functional requirements such as human readability~\cite{zheng2024beyond, gureja2025verification}.
Unless future development workflows allow fully automated AI development without any human code reviews, code readability will remain an important dimension to pursue in AI-generated code.
Addressing this requires both technical innovation (e.g., readability-aware fine-tuning, post-hoc simplification passes) and empirical investigation into what specifically makes LLMs generate overly complicated implementations. 
Until these complexities are addressed, software project teams should treat AI-generated code as requiring extra scrutiny during review, with particular attention to whether simpler implementations exist that achieve the same functionality.

\section{Conclusion}

This study presents the first large-scale empirical investigation of how LLM agent assistants impact real-world software development projects. 
Through a rigorous difference-in-differences design comparing Cursor-adopting repositories with matched control group repositories, complemented by dynamic panel GMM analysis, we provide evidence that challenges both unbridled optimism and categorical pessimism surrounding AI-assisted coding: 
Cursor adoption produces substantial but transient velocity gains alongside persistent increases in technical debt; such technical debt accumulation subsequently dampens future development velocity.
Ultimately, our results suggest a self-reinforcing cycle where initial productivity surges give way to maintenance burdens.

However, several considerations suggest this picture may not be as bleak as it initially appears. 
First, our study captures a snapshot of rapidly evolving technology from mid-2024 to mid-2025, when LLM capabilities, agent designs, and developer practices are improving at unprecedented rates---future tools might be able to address the quality concerns we observed. 
Second, our quality metrics, while well-established in software engineering research, may not fully capture the multi-dimensional nature of code quality in AI-driven development. 
For example, complexity metrics were designed for human-written code; whether they appropriately penalize AI-generated patterns that are mechanically verifiable yet syntactically complex remains an open question. 
Third, the open-source context of our study may amplify both abandonment dynamics and quality concerns relative to enterprise settings with mandatory, dedicated quality assurance processes.

Looking forward, our findings point to clear research and practice directions (Section~\ref{sec:discussion}).
Ultimately, this study demonstrates that realizing the promise of AI-assisted software development requires a holistic understanding of how AI assistance reshapes the fundamental trade-offs between development velocity, code quality, and long-term project sustainability. 
The age of AI coding has arrived---our challenge now is to harness it wisely.

\section*{Data Availability}
\label{sec:data-avail}
We provide a replication package for this paper at:
\begin{quote}
\url{https://doi.org/10.5281/zenodo.18368661}
\end{quote}
The Appendix is available in the latest arXiv version of this paper at: 
\url{https://arxiv.org/abs/2511.04427}.

\begin{acks}
He's, Kästner's, and Miller's work was supported in part by the National Science Foundation (award 2206859).
Miller's work was also supported by the National Science Foundation Graduate Research Fellowship Program under Grant Number DGE214073.
Vasilescu's and Agarwal's work was supported in part by the National Science Foundation (awards 2317168 and 2120323) and research awards from Google and the Digital Infrastructure Fund.
We would like to thank Narayan Ramasubbu and Alexandros Kapravelos for providing valuable methodological feedback at earlier stages of this research.
We would also like to thank all S3C2 Quarterly Meeting attendees for their insightful discussions around the early results of this study.
Finally, we would like to thank Google Cloud for offering research credits to cover BigQuery-based analysis in this research.
\end{acks}

\newpage
\bibliographystyle{ACM-Reference-Format}
\bibliography{references}

\newpage

\appendix

\section{Matching Results}
\label{sec:appendix-matching}

In this section, we present additional details about the propensity score matching process described in Section~\ref{sec:matching}.

Table~\ref{tab:matching} presents the total number of candidate repositories in each major Cursor adoption cohort and the AUCs and McFadden's Pseudo R$^2$s from fitting each logistic regression model in that cohort with 10,000 sampled candidates.
In general, the logistic regression models achieve very high goodness-of-fit (0.8281 to 0.9144 AUC), but only explain a relatively limited amount of variance in the outcome (14.12\% to 27.22\%).
This indicates that while the models can distinguish Cursor adoptions relatively well, the measured covariates alone do not fully explain the variations behind each adoption case.
We view this as an expected result, as the Cursor adoption decisions are probably not driven by any observable information in GitHub; the latter may only latently and partially capture it to help us create a reasonably comparable control group.

\begin{table}
\centering
\small
\caption{Model fitting summary of the logistic regression models used for matching in each Cursor adoption cohort. }
\label{tab:matching}
\begin{tabular}{lrrr}
\toprule
Cohort & Candid. Repos & AUC & McFadden's Pseudo R$^2$ \\
\midrule
202408 & 807,608 & 0.8506 & 0.1412 \\
202409 & 804,740 & 0.9137 & 0.2336 \\
202410 & 818,160 & 0.8969 & 0.2204 \\
202411 & 796,312 & 0.9144 & 0.2722 \\
202412 & 794,186 & 0.8907 & 0.2101 \\
202501 & 776,547 & 0.8702 & 0.2123 \\
202502 & 762,968 & 0.8716 & 0.2271 \\
202503 & 792,382 & 0.8281 & 0.1939 \\
\bottomrule
\end{tabular}
\end{table}

\begin{figure}
     \centering
     \includegraphics[width=0.9\linewidth]{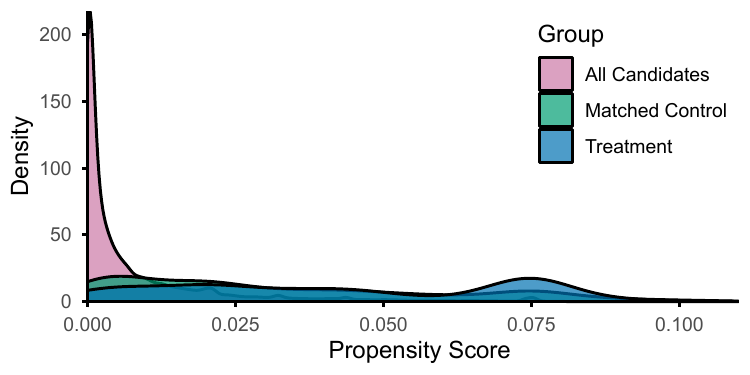}
      \caption{The distribution of propensity scores between all candidate repositories, matched control group repositories, and the \numtreated Cursor-adopting repositories, showing that the matched control group has highly similar conditional treatment probabilities compared to the treatment group.}
     \label{fig:propensity-score}
\end{figure}

To assess whether the treatment and control group repositories are reasonably comparable, we plot the distribution of propensity scores (Figure~\ref{fig:propensity-score}) and conduct balance checks on observable covariates before each adoption (Table~\ref{tab:balance-check}).
While the propensity score plot (Figure~\ref{fig:propensity-score}) shows that the treatment and control group repositories have highly similar conditional treatment probabilities, the matched control group is slightly skewed toward lower propensity scores.
Furthermore, the balance checks (Figure~\ref{tab:balance-check}) indicate acceptable but imperfect matching: While all metrics show acceptable balance between the treatment and control group in terms of normalized differences ($|\text{Norm. Diff}| < 0.25$), there are still observable mean differences between the two groups.
In other words, it indicates that our matching is imperfect and our study setting deviates from perfect randomized controlled experiments.
This motivates us to adopt a difference-in-difference design for causal inference instead of simply comparing outcomes between the two groups---a DiD design with two-way fixed effects and time-varying covariates is more suited for a quasi-experimental setting like ours.

\begin{table}
    \centering
    \small
    \caption{Balance statistics: treatment versus matched control pre-adoption. Normalized difference is defined as $(\bar X_{t} - \bar X_{c})/{\sqrt{(S_t+S_c) / 2}}$, where $\bar X_{t}$/$\bar X_c$ and $S_{t}$/$S_c$ stands for the mean and variance in the treatment/control group, respectively. Follwing balance check conventions~\cite{austin2009balance, stuart2013prognostic}, we consider $|\text{Norm. Diff}| < 0.25$ as acceptable balance and  $< 0.1$ as good balance.
    Note that the treatment means here are different from the treatment means in Table~\ref{tab:repo-stats} because the latter reflect the metrics \emph{at the time of data analysis} while the Table here reflects the same metrics \emph{at the time of Cursor adoption}.}
    \label{tab:balance-check}
    \begin{tabular}{lrrr}
    \toprule
    Metrics & Treatment Mean & Control Mean & Norm. Diff \\
    \midrule
    Age (in days)       & 496.07   & 681.38  & $-$0.207 \\
    Comments     & 2870.82  & 564.84  & 0.147 \\
    Forks        & 196.07   & 64.27   & 0.079 \\
    Issues       & 524.07   & 103.95  & 0.154 \\
    Pull Requests        & 1075.58  & 266.03  & 0.158 \\
    Releases     & 27.51    & 25.44   & 0.006 \\
    Stars        & 1056.65  & 334.62  & 0.130 \\
    Total Events   & 10103.25 & 2443.39 & 0.171 \\
    Users Involved & 1247.79  & 414.90  & 0.137 \\
    \bottomrule
    \end{tabular}
\end{table}

\section{Alternative DiD Estimators}
\label{sec:appendix-alternative-estimators}

Recall from Section~\ref{sec:modeling} that several alternative DiD estimators are available for estimating the average treatment effect on treated $ATT$ and the ``horizon-average'' treatment effect $ATT_h$.
In this section, we provide a brief introduction to the two other widely popular alternative estimators, namely the two-way fixed effects (TWFE) estimator and the ~\citet{callaway2021difference} estimator.
Then, we compare their estimation results with results from the~\citet{borusyak2024revisiting} estimator presented in the main paper.

\subsection{The Two-Way Fixed Effects Estimator}
\label{sec:twfe}

The TWFE estimator estimates $ATT$ from the $\hat\beta$ parameter in the following ordinary least squares regression (OLS) with per-repository ($\mu_i$) and per-period ($\lambda_t$) fixed effects (thus ``two-way''), \emph{on all available observations} with $D_{it} = 1$ for treated and $D_{it}=0$ otherwise:

\begin{equation}
\label{eq:base-did}
    Y_{it} = \hat\mu_i + \hat\lambda_t + \hat\beta D_{it} + \hat\Gamma'Z_{it} + \epsilon_{it}
\end{equation}

For $ATT_h$ and pre-trend test parameters (Equation~\ref{eq:parallel-trend}), the TWFE estimator typically estimates one single OLS regression in the following form, \emph{on all available observations}:
\begin{equation}
\label{eq:panel-event}
    Y_{it}=\hat{\mu}_i + \hat{\lambda}_t + \hat{\Gamma}' Z_{it} + \sum_{h=-k, h\ne -1}^{j}\hat\tau_h\mathbf{1}[t=E_i + h] + \epsilon_{it}
\end{equation}
Here, $\hat\tau_h$ for $h<0$ represents ``placebo'' pre-treatment effect estimates as in Equation~\ref{eq:parallel-trend} and $\hat\tau_h$ for $h\ge0$ represents the intended post-treatment $ATT_h$ estimates.
$h=-1$ is intentionally omitted from the regression to serve as the counterfactual baseline.

While the early econometric literature extensively uses the two-way fixed effect (TWFE) estimator for DiD studies, it is important to note that the OLS-estimated $\hat\beta$s (Equation~\ref{eq:base-did}) and $\hat\tau_h$s (Equation~\ref{eq:panel-event}) do not really estimate the $ATT$ and $ATT_h$ defined in Section~\ref{sec:matching}.
Instead, under both the parallel trend assumption and the treatment effect homogeneity assumption (i.e., the treatment effect does not vary over time and adoption cohorts), a TWFE estimator equals a variance weighted average of all possible two-group/two-period DiD estimators in the data~\cite{goodman2021difference}.
However, the treatment effect homogeneity assumption is a strong assumption that is likely violated in the staggered adoption setting: For example, it is reasonable to anticipate that later Cursor adoption cohorts may have stronger adoption effects as the tooling and model capabilities are rapidly advancing.
If this assumption is violated, it may lead to ``forbidden comparisons'' and negative weighting in the TWFE estimator, biasing the estimated parameters and jeopardizing the validity of causal claims based on TWFE estimators~\cite{de2020two, athey2022design}.

\subsection{The~\citet{callaway2021difference} Estimator}
\label{sec:callaway}

The \citet{callaway2021difference} estimator takes a fundamentally different approach from both TWFE and the \citet{borusyak2024revisiting} imputation estimator. 
Rather than estimating regression models on all adoption cohorts, it first estimates \emph{group-time average treatment effects} $ATT(g,t)$---the average treatment effect for cohort $g$ (repositories adopting in the same period) at time $t$---before aggregating $ATT(g,t)$ into summary $ATT$ and $ATT_h$ measures.

For a group of repositories $g$ that adopt Cursor at time period $g$, the average treatment effect at calendar time $t$ is identified as:
\begin{equation}
\label{eq:cs-att-gt}
ATT(g, t) = \mathbb{E}[Y_t - Y_{g-1} \mid G_g=1] - \mathbb{E}[Y_t - Y_{g-1} \mid C]
\end{equation}
where $Y_t - Y_{g-1}$ represents the evolution of the outcome from the period prior to treatment ($g-1$) to the current period $t$. 
The researcher may choose ``never-treated'' repositories or ``not-yet-treated'' observations as the control group $C$; we choose the latter to align with the other two estimators.
This formulation explicitly avoids using already-treated units as controls, eliminating the ``forbidden comparisons'' that can bias TWFE estimates~\cite{goodman2021difference}.

With $ATT(g, t)$ estimations, \begin{equation}
\label{eq:cs-att}
ATT = \sum_{g} \sum_{t \geq g} w_{g,t} \cdot ATT(g, t)
\end{equation}
where $w_{g,t}$ are weights proportional to the size of each group for all post-treatment observations (i.e., $t \geq g$). $ATT_h$ is given by:
\begin{equation}
\label{eq:cs-agg}
ATT_h  = \sum_{g} \sum_{t} w_{g,t}^h \cdot ATT(g, t)
\end{equation}
where $w_{g,t}^h$ are weights proportional to the size of each group satisfying $t - g = h$. 
As with the other two estimators, this aggregation allows us to test the parallel trends assumption (where $h < 0$) and estimate ``horizon-average'' treatment effects (where $h \geq 0$).

The \citet{callaway2021difference} framework provides multiple approaches to estimate $ATT(g, t)$. 
In our study, we use the \emph{outcome regression} estimator with covariate adjustment.
Specifically, for each group-time pair $(g, t)$, we first estimate an outcome regression model on the comparison group $C$:
\begin{equation}
\label{eq:cs-outcome-reg}
Y_{it} - Y_{i,g-1} = \alpha_{g,t} + \Gamma'_{g,t} Z_{i,g-1} + \epsilon_{it}, \quad \text{for } i \in C
\end{equation}
where $Z_{i,g-1}$ represents pre-treatment covariates (defined in Section~\ref{sec:covariates}).
This model estimates the expected evolution of outcomes for untreated repositories with similar pre-treatment characteristics.
The estimated $ATT(g,t)$ is then computed as:
\begin{equation}
\label{eq:cs-att-or}
\widehat{ATT}(g, t) = \frac{1}{|G_g|}\sum_{i \in G_g} \left[ (Y_{it} - Y_{i,g-1}) - (\hat{\alpha}_{g,t} + \hat{\Gamma}'_{g,t} Z_{i,g-1}) \right]
\end{equation}
where $G_g$ denotes the set of repositories in cohort $g$, and $(\hat{\alpha}_{g,t} + \hat{\Gamma}'_{g,t} Z_{i,g-1})$ represents the predicted counterfactual outcome change for treated repository $i$ based on its pre-treatment characteristics.
We also allow for one period of anticipation in the estimation, consistent with our treatment of the \citet{borusyak2024revisiting} estimator, where we exclude $h=-1$ from pre-trend tests.

A key distinction of the \citet{callaway2021difference} estimator is that it estimates treatment effects \emph{separately within each cohort} before aggregating. 
While this approach ensures clean identification by avoiding cross-cohort contamination, it can reduce statistical power when individual cohorts are small. 
In our setting, where many adoption cohorts contain fewer than 100 repositories (Figure~\ref{fig:time-cursor-adoption}), this cohort-specific estimation may yield noisier estimates compared to \citet{borusyak2024revisiting}, which leverages all untreated observations to fit a single counterfactual outcome model.

\subsection{Comparing Estimation Results}
\label{sec:estimator-comp}

\begin{table*}
\centering
\small
\caption{The estimated average treatment effects on treated (i.e., $ATT$) post Cursor adoption from different DiD estimators. Similar to Table~\ref{tab:average-te}, all outcome variables are log-transformed, and percentage changes are provided for reference.}
\begin{tabular}{lrrrrrr}
\toprule
& \multicolumn{2}{c}{Borusyak et al.~\cite{borusyak2024revisiting}} & \multicolumn{2}{c}{Two-Way Fixed Effects} & \multicolumn{2}{c}{Callaway and Sant'Anna~\cite{callaway2021difference}} \\
\cmidrule(lr){2-3} \cmidrule(lr){4-5} \cmidrule(lr){6-7}
Outcome & Estimate (Std. Err.) & \% Change & Estimate (Std. Err.) & \% Change & Estimate (Std. Err.) & \% Change \\
\midrule
Commits & 0.0260$^{\textcolor{white}{***}}$ (0.0429) & +2.63\% ($\pm$4.40\%) & 0.1641$^{***}$ (0.0386) & +17.83\% ($\pm$4.55\%) & $-$0.0073$^{\textcolor{white}{**}}$ (0.0695) & $-$0.73\% ($\pm$6.90\%) \\
Lines Added & 0.2514$^{*\textcolor{white}{**}}$ (0.1063) & +28.58\% ($\pm$13.7\%) & 0.6029$^{***}$ (0.0876) & +82.74\% ($\pm$16.0\%) & 0.4292$^{**}$ (0.1536) & +53.60\% ($\pm$23.6\%) \\
Static Analysis Warnings & 0.2644$^{***}$ (0.0511) & +30.26\% ($\pm$6.66\%) & 0.1696$^{***}$ (0.0415) & +18.48\% ($\pm$4.92\%) & $-$0.1108$^{\textcolor{white}{**}}$ (0.1254) & $-$10.49\% ($\pm$11.2\%) \\
Duplicated Lines Density & 0.0679$^{\textcolor{white}{***}}$ (0.0448) & +7.03\% ($\pm$4.79\%) & 0.0160$^{\textcolor{white}{***}}$ (0.0390) & +1.61\% ($\pm$3.96\%) & $-$0.0034$^{\textcolor{white}{**}}$ (0.0785) & $-$0.34\% ($\pm$7.82\%) \\
Code Complexity & 0.3481$^{***}$ (0.0538) & +41.64\% ($\pm$7.62\%) & 0.2314$^{***}$ (0.0446) & +26.04\% ($\pm$5.62\%) & $-$0.0387$^{\textcolor{white}{**}}$ (0.1136) & $-$3.80\% ($\pm$10.9\%) \\
\bottomrule
\multicolumn{7}{r}{\textit{Note:} $^{*}p<0.05$; $^{**}p<0.01$; $^{***}p<0.001$} \\
\end{tabular}
\label{tab:average-te-comparison}
\end{table*}

\begin{figure*}
    \centering
    \vspace{-1mm}
    \includegraphics[width=\linewidth]{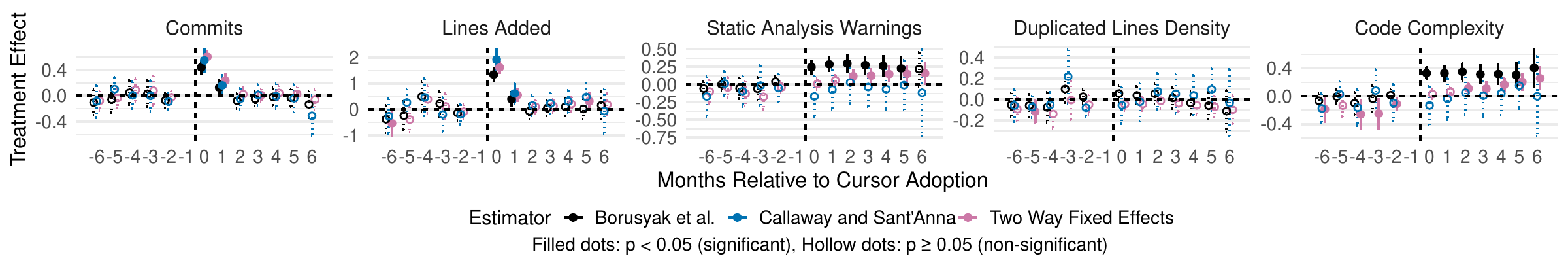}
    \caption{The estimated horizon-average treatment effects ($ATT_h$) from three DiD estimators: \citet{borusyak2024revisiting}, \citet{callaway2021difference}, and two-way fixed effects. The estimates on development velocity outcome are highly consistent and robust across all three estimators, but the estimates on software quality outcomes vary significantly.}
    \label{fig:all-estimators}
\end{figure*}

We summarize the $ATT$ and $ATT_h$ estimations from all three estimators in Table~\ref{tab:average-te-comparison} and Figure~\ref{fig:all-estimators}. 
For pre-trend tests, we use the same heteroskedasticity- and cluster-robust Wald tests~\citep{borusyak2024revisiting} to test the joint null hypothesis that all ``placebo'' pre-treatment effect estimates are equal to zero.
Most models pass the pre-trend test at the 0.05 significance level. 
The one exception is code complexity estimated with TWFE, which shows marginally significant pre-trends visible in Figure~\ref{fig:all-estimators}. This violation likely stems from the ``forbidden comparisons'' inherent to the TWFE estimator~\citep{de2020two}, as both the \citet{borusyak2024revisiting} and \citet{callaway2021difference} estimators show no significant pre-trends for this outcome.

For development velocity outcomes, the three estimators show qualitatively consistent results on the $ATT_h$ estimates (Figure~\ref{fig:all-estimators}).
The differences in $ATT$ estimates stem from the fact that they are averaged differently in the three estimators.
All estimators find positive effects on lines added, with estimates ranging from +28.58\% (\citet{borusyak2024revisiting}) to +82.74\% (TWFE) to +53.60\% (\citet{callaway2021difference}).
The differences in manitudes stem from the fact that they use different weighted averages for the $ATT$ estimates.
While the magnitudes differ, the direction and statistical significance align, providing robust evidence that Cursor adoption increases code output. For commits, \citet{borusyak2024revisiting} and \citet{callaway2021difference} find small, statistically insignificant effects (+2.63\% and $-$0.73\%, respectively), while TWFE estimates a larger, significant effect (+17.83\%). This TWFE inflation likely reflects the bias from ``forbidden comparisons'' discussed in Section~\ref{sec:twfe}.

For code quality outcomes, the estimators diverge more substantially. The \citet{borusyak2024revisiting} and TWFE estimators consistently find significant increases in static analysis warnings (+30.26\% and +18.48\%) and code complexity (+41.64\% and +26.04\%), while the \citet{callaway2021difference} estimator yields negative but statistically insignificant estimates for these same outcomes ($-$10.49\% and $-$3.80\%). 
All three estimators find no significant effects on duplicated lines density after Cursor adoption.

The divergence between \citet{callaway2021difference} and the other estimators warrants careful interpretation, which may be attributed to several methodological differences. 
First, as discussed in Section~\ref{sec:callaway}, the \citet{callaway2021difference} estimator estimates treatment effects separately within each cohort before aggregating. 
In our setting, many adoption cohorts contain fewer than 100 repositories (Figure~\ref{fig:time-cursor-adoption}), meaning cohort-specific estimates rely on limited sample sizes. 
Combined with the inherent noisiness of code quality metrics (e.g., measurement errors in static analysis tools and high variation in code characteristics), this small-cohort structure likely reduces statistical power to detect genuine effects. 
Second, the \citet{callaway2021difference} estimator conditions only on pre-treatment covariates $Z_{i,g-1}$ (Equation~\ref{eq:cs-outcome-reg}), whereas both TWFE and \citet{borusyak2024revisiting} adjust for time-varying covariates $Z_{it}$. 
If time-varying confounders influence code quality outcomes beyond what pre-treatment characteristics capture, this difference in covariate adjustment could contribute to the divergent estimates. 
Still, there is no clear consensus on when time-varying covariates help or hurt causal inference in staggered DiD settings.

We report the \citet{borusyak2024revisiting} estimator in the main paper for three reasons: (1) It avoids the biases of TWFE while maintaining statistical power by pooling untreated observations; (2) it passes pre-trend tests across all outcomes; and (3) the sustained temporal patterns in Figure~\ref{fig:all-estimators} support the causal interpretation. 
We acknowledge that the lack of large cohorts and clear theoretical guidance on the incorporation of time-varying covariates in DiD limits our ability to definitively resolve estimator disagreements for code quality outcomes; the findings around code quality outcomes should be interpreted with appropriate caution.

\section{Full Robustness Check Results}

\begin{table*}[t]
\centering
\small
\caption{Summary statistics for all alternative dataset settings explored in this study. The results from Rust repositories are dropped in Figure~\ref{fig:robustness-checks-all} because of insufficient observations, causing insignificant results and extremely large confidence intervals.}
\begin{tabular}{lrrrr}
\toprule
Setting & \# Treatment Repos & \# Control Repos & \# Total Observations & \# Post-Treatment Observations \\
\midrule
\multicolumn{5}{l}{\textbf{Main Settings}} \\
\quad Original & 801 & 1,172 & 21,699 & 4,481 \\
\quad High Contributor Adoption & 379 & 535 & 8,440 & 1,919 \\
\quad Cursor Configuration Changes & 801 & 1,127 & 18,319 & 1,569 \\
\quad Active Months ($>$0 commits) & 801 & 1,172 & 16,075 & 3,934 \\
\quad Very Active ($\geq$10 commits) & 709 & 814 & 10,134 & 2,721 \\
\quad Cursor and Others & 382 & 583 & 11,322 & 2,440 \\
\quad Only Cursor & 419 & 602 & 10,232 & 2,041 \\
\addlinespace
\multicolumn{5}{l}{\textbf{Adoption Cohort Settings}} \\
\quad Before Agent Release	& 166 & 286 & 4,793 & 1,262 \\
\quad After Agent Release	& 635 & 769 & 15,508 & 3,219 \\
\quad Agent Made Default	& 461 & 570	& 11,748 & 2,199 \\
\addlinespace
\multicolumn{5}{l}{\textbf{Programming Language Settings}} \\
\quad JavaScript/TypeScript & 411 & 422 & 8,870 & 2,279 \\
\quad Python & 121 & 127 & 2,461 & 582 \\
\quad Go & 35 & 57 & 1,147 & 191 \\
\quad Rust & 21 & 41 & 771 & 132 \\
\bottomrule
\end{tabular}
\label{tab:summary-stats-settings}
\end{table*}

\begin{figure*}
    \centering
    \includegraphics[width=\linewidth]{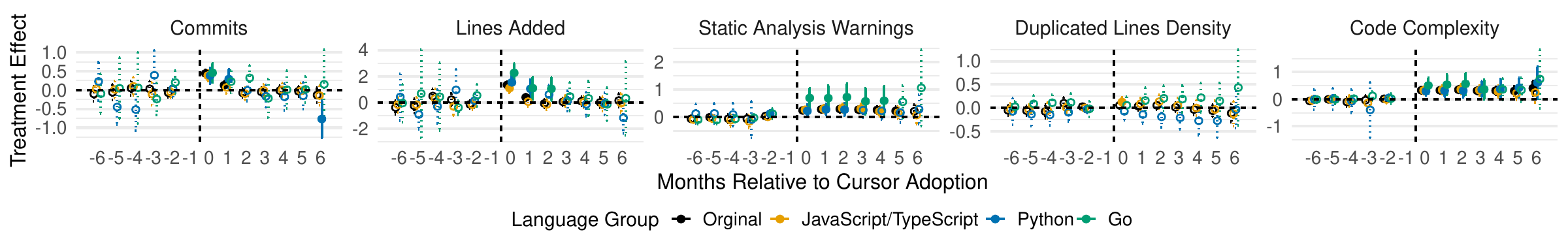}
    \includegraphics[width=\linewidth]{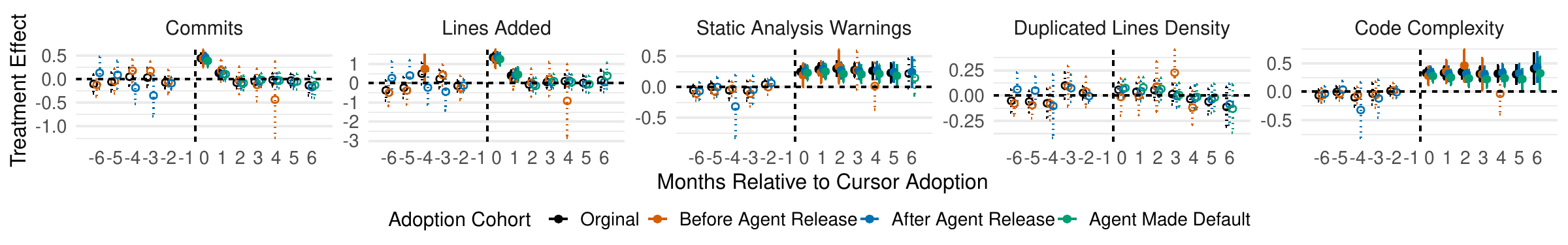}
    \caption{The estimated ``horizon-average'' treatment effects in alternative dataset settings with all five outcome variables.
    \emph{Row 1}: Robustness check across programming language groups with statistically sufficient observations in our dataset (JavaScript/TypeScript, Python, Go), showing that our findings are qualitatively consistent across programming languages.
    \emph{Row 2}: Robustness checks across Cursor adoption cohorts where different Cursor features were available, showing no qualitative difference across cohorts before/after agent release and agent made default in the Composer feature.}
    \label{fig:robustness-checks-all}
\end{figure*}

Following discussions in Section~\ref{sec:robustness}, we present summary statistics in all alternative dataset settings in Table~\ref{tab:summary-stats-settings} and additional robustness check results in Figure~\ref{fig:robustness-checks-all}.
Apart from the settings already discussed in Section~\ref{sec:robustness}, we also conduct experiments across different programming language groups to determine whether our findings are driven by repositories in particular programming languages or are consistent across programming languages.
The results in Figure~\ref{fig:robustness-checks-all}, Row 1 show that our main causal findings are qualitatively consistent across all the major programming languages in our dataset (JavaScript/TypeScript, Python, Go).
Comparing the differences, we observe that the velocity gain is most transient in JavaScript/TypeScript repositories but less so in Python and Go repositories, while the effect on code quality in Go repositories is strongest.
However, it is essential to note that this heterogeneity should not be attributed to the programming languages themselves in our setting. Our intuition is that some form of selection bias (e.g., Go repositories may be larger and more sustained) is driving the heterogeneity we observe here.
Also, our dataset does not sufficiently cover several major programming languages, such as Java, C/C++, and Rust, in which the outcomes of AI coding tool adoption might differ substantially from those in Python and JavaScript projects (the latter have more training data and better LLM performance as of now).
Therefore, we believe that future research is necessary to explore the impact of AI coding tools on other programming languages and the mechanisms underlying these heterogeneities.

Another limitation of our study is that our identification based on Cursor rule files cannot clearly distinguish developers who use the older Cursor Composer feature (with autonomous file editing but without agentic capabilities) from those who use the later Agentic features.
The reality is probably that most repositories in our dataset have a mix of developers using different AI autonomy levels, with most switching to full-fledged agents after it became the default in February 2025.
As another robustness check, we check whether the availability of the agentic feature impacts our main results through different adoption cohorts, \emph{Before Agent Release}, for the adoption cohorts before November 2024, \emph{After Agent Release} for the adoption cohorts after November 2024, and \emph{Agent Made Default} for the adoption cohorts after Feburary 2025.
Note that clear separation of feature availability is impossible in our longitudinal study setting, as it is very likely that early Cursor adopters will also switch to using agents later during our observation period.
The results in Figure~\ref{fig:robustness-checks-all}, Row 2 show no qualitative difference across the three adoption cohorts for the main findings, except that the estimated effects on static analysis warnings and code complexity are slightly weaker.
As with the pprogramminglanguage case, the interpretation of this difference is challenging.
The effect may be because of an increase in model capabilities, merely selection bias, or because most repositories eventually switch to agents.

\section{Analysis of SonarQube Warnings}

\begin{table}
\centering
\caption{The estimated average number of new static analysis warnings introduced per repository per month in each SonarQube warning category pre-/post-Cursor adoption. See Table~\ref{tab:warning-taxonomy} for the definition of each category.}
\label{tab:warnings-by-category}
\small
\begin{tabular}{lrrr}
\toprule
\textbf{Category} & \textbf{Pre Mean} & \textbf{Post Mean} & \textbf{Change} \\
\midrule
Naming Conventions    & 12.15 & 33.48 & +21.32 \textcolor{red}{$\uparrow$} \\
Code Hygiene          &  6.58 & 22.72 & +16.15 \textcolor{red}{$\uparrow$} \\
Code Complexity       &  7.59 & 22.92 & +15.34 \textcolor{red}{$\uparrow$} \\
Code Style            & 14.51 & 29.27 & +14.76 \textcolor{red}{$\uparrow$} \\
Data Science          &  2.10 & 13.39 & +11.29 \textcolor{red}{$\uparrow$} \\
React Patterns        &  9.24 & 18.75 & +9.52 \textcolor{red}{$\uparrow$} \\
Type Safety           &  7.61 & 16.04 & +8.43 \textcolor{red}{$\uparrow$} \\
CSS Issues            &  3.41 &  9.09 & +5.68 \textcolor{red}{$\uparrow$} \\
OOP/Design            &  4.53 &  7.90 & +3.37 \textcolor{red}{$\uparrow$} \\
Regex Issues          &  2.37 &  5.08 & +2.71 \textcolor{red}{$\uparrow$} \\
Infrastructure        &  2.62 &  5.10 & +2.48 \textcolor{red}{$\uparrow$} \\
Logic Error           &  7.57 &  9.66 & +2.09 \textcolor{red}{$\uparrow$} \\
Security              &  1.78 &  3.76 & +1.98 \textcolor{red}{$\uparrow$} \\
Empty/Incomplete Code &  5.37 &  6.98 & +1.61 \textcolor{red}{$\uparrow$} \\
Error Handling        &  5.28 &  6.44 & +1.16 \textcolor{red}{$\uparrow$} \\
Accessibility         & 11.99 & 12.77 & +0.78 \textcolor{red}{$\uparrow$} \\
HTML Structure        &  1.89 &  1.63 & $-$0.26 \textcolor{green}{$\downarrow$} \\
Resource Management   &  3.00 &  2.60 & $-$0.40 \textcolor{green}{$\downarrow$} \\
Concurrency           &  5.00 &  3.05 & $-$1.95 \textcolor{green}{$\downarrow$} \\
API Usage             & 17.77 & 13.42 & $-$4.35 \textcolor{green}{$\downarrow$} \\
\bottomrule
\end{tabular}
\end{table}

To peek into what is actually driving the increase in static analysis warnings, we collect a sample of SonarQube warnings pre- and post-Cursor adoption for the treated repositories. 
It is merely a convenience sample, as an architectural limitation in our SonarQube analysis pipeline prevents us from precisely collecting all warnings and tracking each warning to the version that introduced it.
Thus, we only use this sample for descriptive explorations rather than for causal inference (e.g., using a DiD design as in the main paper).

In total, we collected 195,010 warnings generated from 933 SonarQube analysis rules.
The first author of the paper iteratively works with Claude Opus 4.5 to generate a taxonomy of these rules with 20 categories (Table~\ref{tab:warning-taxonomy}).
Using this taxonomy, we estimate the average number of new static analysis warnings introduced per repository per month in each SonarQube warning category pre-/post-Cursor adoption (Table~\ref{tab:warnings-by-category}) using the available warnings and months.
We only focus on newly introduced warnings in each month where data is still available as of January 2026, as SonarQube may routinely clean up resolved warnings in older project versions. 

Table~\ref{tab:warnings-by-category} reveals that 16 out of the 20 warning categories increased post-Cursor adoption, while only four decreased.
The largest increases occurred in Naming Conventions (+21.32), Code Hygiene (+16.15), Code Complexity (+15.34), and Code Style (+14.76), all of which indicates that the adoption of Cursor---and the rapid development velocity associated with it---may lead to violation of common coding conventions, accumulation of artifacts (TODOs, commented out code, unused variables), and more complex code (e.g., deeply nested functions).
We also observe increases in violation of domain-specific best practices in Data Science (+11.29), React (+9.52), Type Safety (+8.43), CSS (+5.68), etc.
These violations may come from current LLMs being trained on low-quality code or developers heavily vibe-coding and not rigorously reviewing AI-generated code, but future research is necessary to explore the true causes behind them.
Interestingly, Logic Errors and Security problems---actual bugs that are both obvious enough to be detectable by static analysis and potentially dangerous---also increased modestly (+2.09 and +1.98).
The few categories that decreased include API Usage ($-$4.35), Concurrency ($-$1.95), Resource Management ($-$0.40), and HTML Structure ($-$0.26), potentially indicating that AI models trained on recent code may suggest more modern API alternatives and handle certain asynchronous patterns more effectively.

Overall, while the majority of the new static analysis warnings introduced after Cursor adoption are style and maintainability issues, we also observe non-negligible increases in warning categories signaling bad coding practices (e.g., type safety) and critical bugs (e.g., security).
These findings strengthen our recommendation in Section~\ref{sec:practical-implications}, that high-velocity AI-powered development generally introduces, rather than resolves, code quality issues, and quality assurance needs to scale with this AI-era velocity. 

\begin{table*}
\centering
\caption{A taxonomy of static analysis warnings generated by SonarQube in our dataset.}
\label{tab:warning-taxonomy}
\small
\begin{tabular}{lp{7cm}p{7.2cm}@{}}
\toprule
\textbf{Category} & \textbf{Description} & \textbf{Example Rules} \\
\midrule

Code Hygiene & 
Leftover artifacts from development, including commented-out code, TODO/FIXME comments, unused variables, etc. &
\texttt{S125} (commented code), \texttt{S1135} (TODO comments), \texttt{S1128} (unused imports), \texttt{S1481} (unused variables) \\
\addlinespace

Logic Error & 
Bugs in code logic, such as incorrect comparisons, unreachable code, infinite loops, and misuse of return values. &
\texttt{S2871} (sort with ill-defined comparator), \texttt{S1764} (identical sub-expressions), \texttt{S1763} (unreachable code) \\ 
\addlinespace

Code Complexity & 
Structural issues affecting maintainability, including high cognitive complexity, deep nesting, and excessive parameters. &
\texttt{S3776} (cognitive complexity), \texttt{S2004} (deep nesting), \texttt{S3358} (nested ternary), \texttt{S107} (too many parameters) \\
\addlinespace

Accessibility & 
Web accessibility violations, including missing alt text, keyboard navigation issues, and improper ARIA usage. &
\texttt{S1082} (missing keyboard handler), \texttt{S5256} (missing table headers), \texttt{S6848} (non-native interactive elements) \\
\addlinespace

React Patterns & 
React-specific anti-patterns, including incorrect key usage, hook violations, and state mutation issues. &
\texttt{S6479} (array index as key), \texttt{S6440} (hook rules violation), \texttt{S6756} (direct state mutation) \\
\addlinespace

OOP/Design &
Object-oriented design issues, including encapsulation violations, improper inheritance, and design pattern misuses. &
\texttt{S1118} (add private constructor), \texttt{S1104} (public mutable field), \texttt{S2160} (override equals) \\
\addlinespace

Type Safety & 
Type system issues primarily in TypeScript, including unnecessary assertions, improper generics, and enum problems. &
\texttt{S4325} (unnecessary type assertion), \texttt{S6759} (props not readonly), \texttt{S4621} (duplicate union types) \\
\addlinespace

API Usage & 
Use of deprecated or outdated APIs and methods that have newer alternatives. &
\texttt{S1874} (deprecated API usage), \texttt{S6653} (use \texttt{Object.hasOwn}), \texttt{S6654} (\texttt{\_\_proto\_\_} deprecated) \\
\addlinespace

Error Handling & 
Incomplete or improper exception and error handling, including empty catch blocks and generic exceptions. &
\texttt{S108} (empty block), \texttt{S112} (generic exception), \texttt{S1143} (return in finally), \texttt{S2737} (empty catch) \\
\addlinespace

Security & 
Security vulnerabilities, including hardcoded secrets, insecure configurations, and unsafe operations. &
\texttt{S6437} (secrets in image), \texttt{S2819} (\Code{postMessage} origin), \texttt{S5542} (insecure cipher) \\
\addlinespace

Naming Conventions & 
Violations of language-specific naming conventions for classes, methods, variables, and constants. &
\texttt{S100} (method naming), \texttt{S101} (class naming), \texttt{S117} (variable naming) \\
\addlinespace

Regex Issues & 
Regular expression problems, including excessive complexity, empty matches, and inefficient patterns. &
\texttt{S5843} (regex complexity), \texttt{S5869} (duplicate \texttt{char} class), \texttt{S6019} (reluctant quantifier) \\
\addlinespace

Empty/Incomplete & 
Missing implementations, including empty methods, empty classes, and stub code without logic. &
\texttt{S1186} (empty method), \texttt{S2094} (empty class), \texttt{S4658} (empty CSS block) \\
\addlinespace

Concurrency & 
Threading and asynchronous programming issues, including race conditions and improper synchronization. &
\texttt{S2168} (double-checked locking), \texttt{S4123} (redundant await), \texttt{S2696} (non-static in static) \\
\addlinespace

Code Style & 
Formatting and stylistic preferences, including boolean literals, loop styles, and modern syntax usage. &
\texttt{S1125} (boolean literal), \texttt{S1301} (\texttt{switch} vs \texttt{if}), \texttt{S6582} (optional chaining) \\
\addlinespace

Data Science & 
Python ML/data science specific issues, including random seeds, deprecated NumPy patterns, and DataFrame operations. &
\texttt{S6709} (missing random seed), \texttt{S6734} (\texttt{inplace=True}), \texttt{S6730} (deprecated numpy type) \\
\addlinespace

Infrastructure & 
Docker, Kubernetes, and infrastructure-as-code issues, including image versioning and resource specifications. &
\texttt{S6596} (unversioned image), \texttt{S6597} (\texttt{cd} vs \texttt{WORKDIR}), \texttt{S6873} (missing memory request) \\
\addlinespace

Resource Management & 
Resource lifecycle management issues, including unclosed streams and improper cleanup. &
\texttt{S2093} (try-with-resources), \texttt{S2095} (unclosed resource), \texttt{S4042} (\texttt{Files.delete}) \\
\addlinespace

CSS Issues & 
Stylesheet-specific problems, including invalid properties, unknown units, and duplicate selectors. &
\texttt{S4654} (unknown property), \texttt{S4653} (unknown unit), \texttt{S4666} (duplicate selector) \\
\addlinespace

HTML Structure & 
HTML document structure issues, including missing doctype and improper tag nesting. &
\texttt{DoctypePresenceCheck}, \texttt{MetaRefreshCheck}, \texttt{S4645} (unclosed script) \\

\bottomrule
\end{tabular}
\end{table*}

\end{document}